\documentclass[12pt]{article}
\usepackage[font=small]{caption}
\usepackage[english]{babel}
\usepackage{float}
\usepackage{graphicx}
\usepackage{amsmath}
\usepackage{amssymb, amsthm}
\usepackage{array}
\usepackage{verbatim}
\usepackage[T1]{fontenc}
\usepackage[round]{natbib}

\DeclareMathOperator{\logit}{logit}
\DeclareMathOperator{\E}{\mathsf{E}}   
\DeclareMathOperator{\mubase}{\mu_{\textrm{base}}}
\DeclareMathOperator{\muback}{\mu_{\textrm{back}}}

\begin{document}

\newtheorem{theorem}{Theorem}[section]
\newtheorem{proposition}[theorem]{Proposition}
\newtheorem{example}{Example}[section]

\pagenumbering{arabic}

\title{Stochastic frailty models for modeling \\ and forecasting mortality}
\date{October, 2014}
\author{S{\o}ren Fiig Jarner\footnote{The Danish Labour Market Supplementary Pension Fund (ATP), Denmark; Department of Mathematical Sciences, University of Copenhagen, Denmark}}

\bigskip\bigskip
\maketitle
\bigskip

\noindent{\bf Abstract.}
In many countries life expectancy gains have been substantially higher than predicted by even recent forecasts. This
is primarily due to increasing rates of improvement in old-age mortality not captured by existing models. In this paper
we show how the concept of frailty can be used to model both changing rates of improvement and the deceleration of mortality at old ages, also seen in data.

We present a "fragilization" method by which frailty can be added to standard mortality models. The
aim is to improve the modeling and forecasting of old-age mortality while preserving the structure
of the original model and the underlying stochastic processes. Estimation is based on a general
pseudo-likelihood approach which allows the use of essentially any frailty distribution and mortality model.
We also consider a class of generalized stochastic frailty models with both frailty and non-frailty terms, and we describe how
these models can be estimated by the EM-algorithm.

The method is applied to the Lee-Carter model and a parametric time-series model.
For both applications the effect of adding frailty is illustrated with mortality data for US males.

\bigskip


\thispagestyle{empty}
\newpage

\section{Introduction}  \label{sec:intro}
Mortality rates have been steadily declining in most of the industrialized world throughout the 20th century and there are no signs
of improvements decelerating. On the contrary, populations where mortality rates are already very low still experience rates
of improvement of the same, or even higher, magnitude than historically.

A great deal of models for modeling and projecting mortality rates have been proposed in recent years. However,
the model formulated by \citet{leecar92} is still the most widely used.
The Lee-Carter model describes the age-specific death rates (ASDRs) by the log-bilinear relation
\begin{equation} \label{eq:LC}
  \log \mu(t,x) = a_x + b_x k_t,
\end{equation}
where $a$ and $b$ are age-specific parameters and $k$ is a time-varying index. Projections are made by estimating the parameters
of the model from observed data and extrapolate the time index $k$ by standard time series methods. Typically $k$
is modeled as a random walk with drift. The combination of a linear forecast of $k$ and structure (\ref{eq:LC}) implies
that ASDRs are forecasted to improve by the same age-specific factor each year.

The Lee-Carter model has gained widespread popularity due to its simplicity and ease of interpretation. Essentially, ASDRs are forecasted
to improve by the average rate of improvement over the estimation period. However, the assumption that rates of improvement are constant
over time is not generally satisfied. Indeed, the mortality experience of several countries has shown increasing mortality improvement
rates over time, in particular for older age groups, see e.g.\ \citet{leemil01,booetal02,renhab03,bon05}. Consequently, Lee-Carter
forecasts have in many situations underestimated the gains in old-age mortality.

\begin{figure}[h]
\begin{center}
\includegraphics[height=7.5cm]{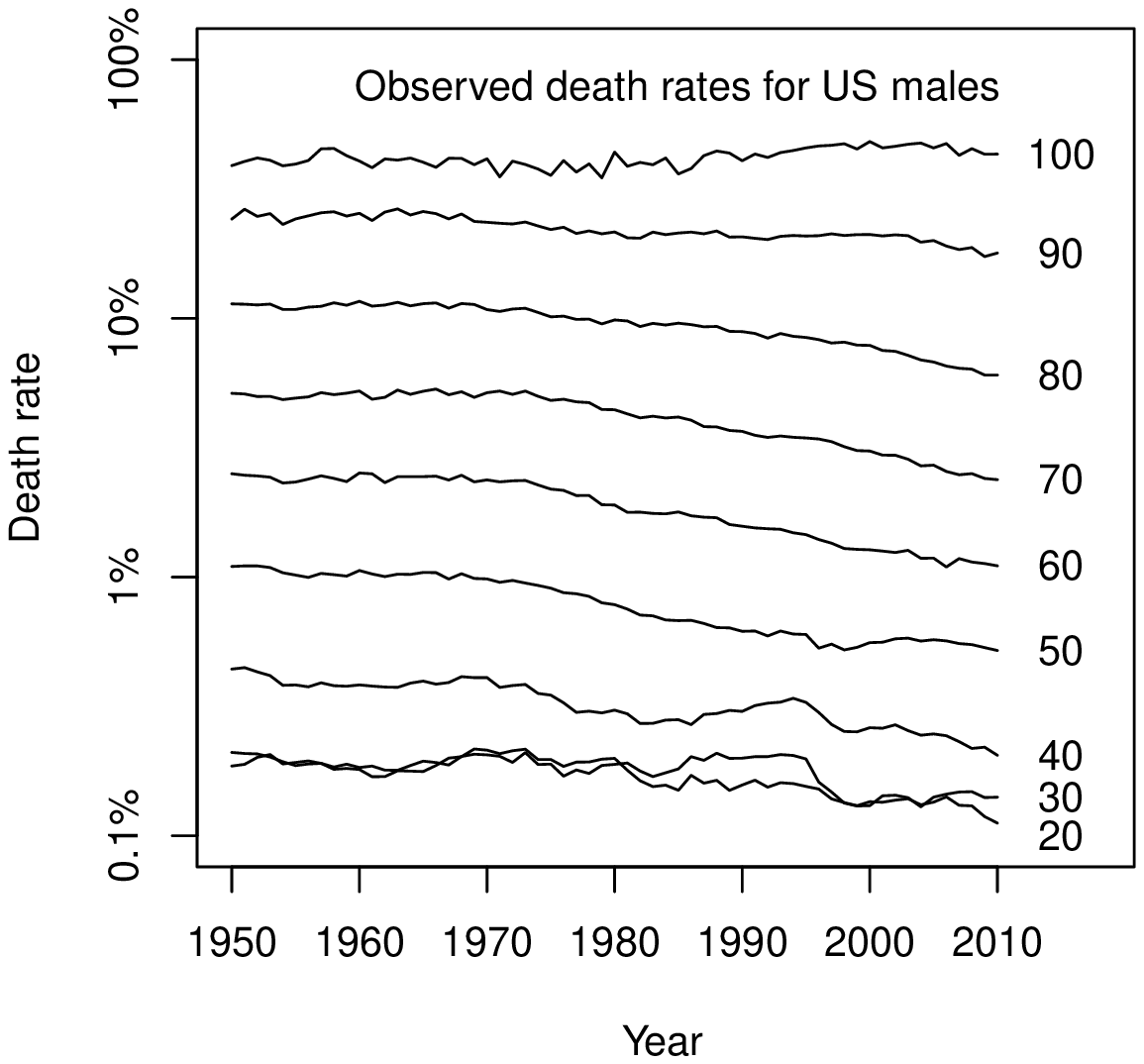}
 \hfill
\includegraphics[height=7.5cm]{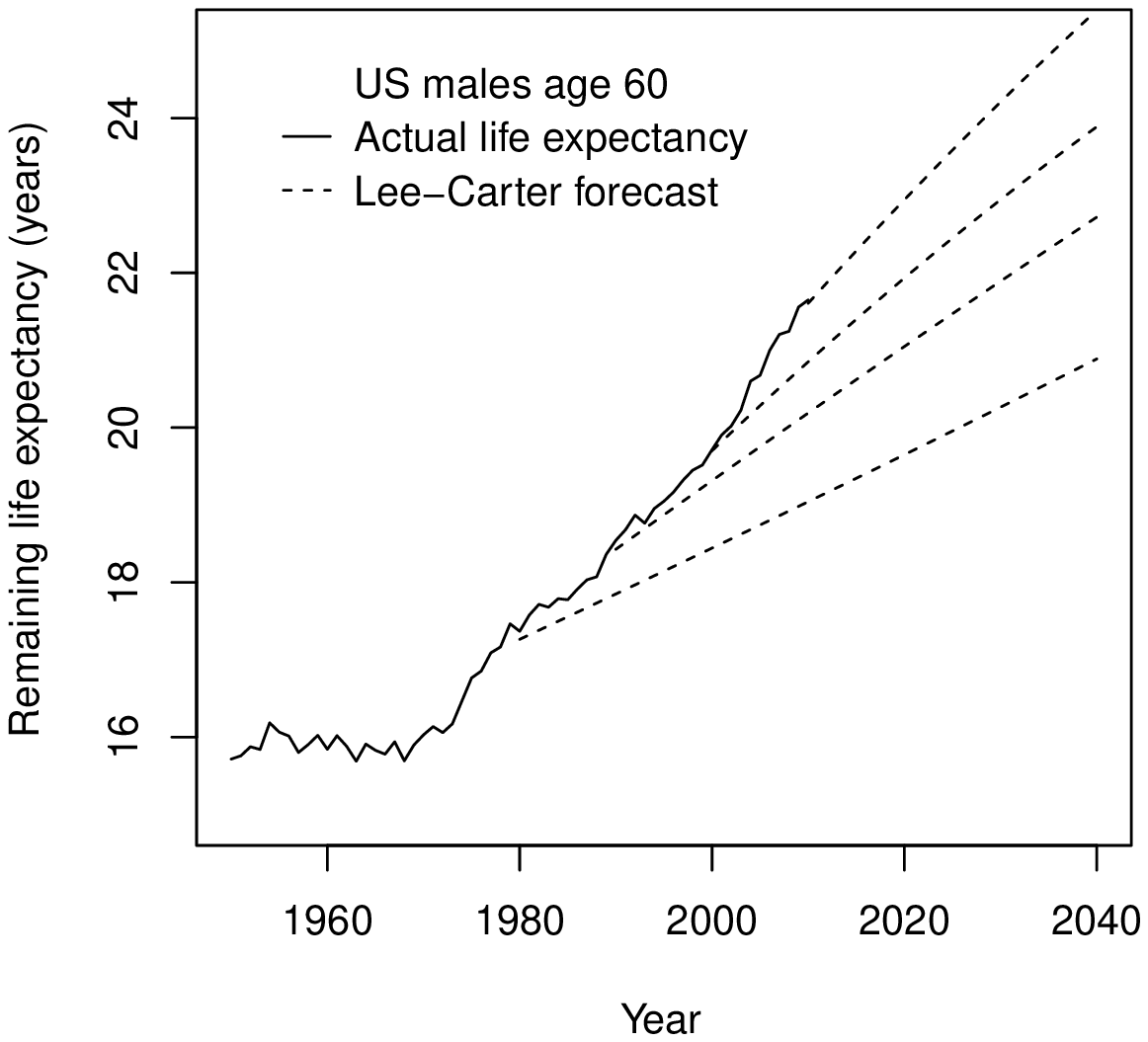}
\end{center}
\vspace*{-5mm}
\caption{Left panel shows observed death rates for US males from 1950 to 2010 for ages 20, 30, $\ldots$, 100 years.
Right panel shows remaining period life expectancy of US males age 60. The solid line is the actual life
expectancy and the dashed lines are forecasts based on the Lee-Carter model with estimation periods 1950--1980, 1960--1990, 1970--2000 and 1980--2010.}
\label{fig:LCintro}
\end{figure}

Left panel of Figure~\ref{fig:LCintro} shows ASDRs for US males from 1950 to 2010 based on data from the Human Mortality Database (www.mortality.org).
The death rates are plotted on a logarithmic scale and increasing rates of improvement can be observed by the concavity, rather than linearity,
of the age-specific curves. The consequence for Lee-Carter forecasts is exemplified in the right panel of Figure~\ref{fig:LCintro} which
shows actual and Lee-Carter forecasted period remaining life expectancy of 60-year old US males.\footnote{The period (remaining) life expectancy
is a summary measure for the mortality profile of a population in a given year. It is calculated on the basis of the age-specific mortality rates for that year.
Assuming death rates continue to decline the life expectancy experienced by a given cohort will be higher. This quantity is termed
the cohort life expectancy.} Due to increasing rates
of improvement the forecasts increase when a later estimation period is used. Still, all historic forecasts are below the actual life expectancy experience.

In the light of changing rates of mortality improvement it has been suggested to fit the Lee-Carter model to
shorter periods of data for which death rates comply better with the log-linearity assumption, see e.g.~\citet{leemil01,tuletal00,booetal02}.
In an environment of increasing rates of improvement the use of shorter, more recent data periods can indeed result in better
forecasts. However, future increases in old-age mortality improvement rates --- beyond what
has been seen historically --- cannot be captured by this method.

\subsection{Stochastic frailty models}
In this paper we investigate the use of frailty theory to address the pattern of changing rates of improvement. Frailty theory
rests on the assumption that cohorts are heterogeneous and that some people are more susceptible to death (frail) than others.
Frail individuals tend to die sooner than stronger individuals leading to old cohorts being dominated by low mortality individuals.
In effect, the selection mechanism causes the cohort intensity to "slow down", i.e.\ to increase less rapidly than the individual intensity.
This has been used to explain the logistic form of old-age mortality observed in data, see e.g.\ \citet{tha99}.

In the model introduced by \citet{vauetal79} frailty enters as an unobservable, positive random quantify $Z$ acting multiplicatively
on an underlying baseline mortality intensity. The observable (aggregate) mortality intensity takes the form
\begin{equation}   \label{eq:mulfrail}
   \mu(t,x) = \E[Z |t,x] \mubase(t,x),
\end{equation}
where $\E[Z |t,x]$ denotes mean frailty at time $t$ for age $x$, and $\mubase(t,x)$ is the underlying baseline
intensity for an individual with frailty one. We assume without loss of generality that mean frailty at birth is one.
For young ages with low mortality mean frailty is close to one,
while for old ages mean frailty gradually decreases as selection takes effect.

Frailty theory offers an explantation to the observed change in improvement rates of old-age mortality and it suggests that we
might expect to see higher rates of improvements in the future.
As the probability of attaining a given (old) age increases over time the selection effect weakens and
mean frailty increases towards one. This increase partly offsets the underlying improvements in
individual baseline intensity, cf.\ (\ref{eq:mulfrail}). Initially when selection is high old-age mortality rates will improve slowly (or even increase), but over time
as selection weakens improvements in old-age mortality will gradually increase towards baseline improvement rates.

In this paper we will use structure (\ref{eq:mulfrail}) as the basis for mortality modeling.
We will show how a given model for aggregate mortality, e.g.\ the Lee-Carter model, can be "fragilized" by using it
as a model for the underlying individual mortality instead.
This procedure will enable us to model logistic-type old-age mortality and changing rates of improvement
while preserving the structure of the original model and the underlying time-series dynamics. As an example, a "fragilized" Lee-Carter model
takes the form
\begin{equation}   \label{eq:LCfrail}
   \mu(t,x) = \E[Z |t,x] \exp[a_x + b_x k_t].
\end{equation}
In Section~\ref{sec:app2} we will compare this model to the ordinary Lee-Carter model.

We show how to perform maximum likelihood estimation and how to forecast in models of form (\ref{eq:mulfrail}).
Estimation will be based on a Poisson pseudo-likelihood which allows for general specifications of the frailty distribution,
e.g.\ the generalized stable laws of \citet{hou86}.
We also generalize (\ref{eq:mulfrail}) to include an additive (non-frailty) term and show how models of this
form can be estimated by the EM-algorithm. Methods for joint maximum likelihood estimation of the frailty distribution
and the baseline intensity are derived based on the (conditional) likelihood for fixed frailty distribution.

In contrast to the typical use of frailty theory in mortality modeling our approach does not rely
on a specific parametric model. Indeed, we show how essentially any combination of frailty distribution and parametric or semi-parametric
model for baseline intensity can be estimated. The only requirement is that the underlying
baseline mortality model can be estimated (by maximum likelihood) and that the Laplace transform of the frailty distribution
is known.

Frailty theory is well-established in biostatics and survival analysis and several monographs are devoted to the topic,
e.g.\ \citet{wie10,han11}. In demographic and actuarial science frailty models are also known as heterogeneity models. They have been used in mortality
modeling to fit the logistic form of old-age mortality, see e.g.\ \citet{wanbro98,buthab04,oli06,caibladow06,spretal13},
and to allow for overdispersion in mortality data, cf.\ \citet{lietal09}.

The present use of frailty theory is closest in spirit to \citet{jarkry11} in which a deterministic frailty model
is used to describe the trend in international mortality. However, the methods developed in this paper
greatly extend that work by incorporating frailty into any standard stochastic mortality model.

\subsection{Outline}
The rest of the paper is organized as follows. In Section~\ref{sec:frail} we give a brief introduction to frailty theory to establish the
fundamental relation between cohort and individual mortality on which our approach relies. Section~\ref{sec:SFM} is the main theoretical chapter
in which stochastic frailty models are formalized and estimation and forecasting discussed. This is followed in Section~\ref{sec:app} by applications
to a parametric time-series model and the Lee-Carter model. Section~\ref{sec:genSFM} contains a generalization of stochastic frailty models and explains how they can be estimated by the EM-algorithm. Finally, Section~\ref{sec:remarks} offers some concluding remarks.

\section{Frailty theory} \label{sec:frail}
In this section we go through the basic theory and establish relations to be used in the following.
For ease of exposition we consider only a single birth cohort in a continuous-time model. Later on we will add a time parameter.
As our starting point we take the frailty model of \citet{vauetal79} in which frailty is modeled as an unobservable, non-negative quantity $Z$
acting multiplicatively on an underlying, baseline (mortality) intensity. The intensity, or death rate, of an
individual of age $x$ with {\em frailty} $z$ is given by
\begin{equation}
   \mu(x;z) = z \mubase(x),
\end{equation}
where $\mu_{\textrm{base}}$ is the baseline intensity describing the age effect. The corresponding survival function, i.e.\ the probability
of the person surviving to age $x$, is
\begin{equation}
   S(x;z) = \exp\left(-\int_0^x \mu(u;z)du \right) = \exp\left(-z I(x) \right),
\end{equation}
where $I$ denotes the integrated baseline intensity,
\begin{equation}
   I(x) = \int_0^x \mubase(u)du.
\end{equation}

The survival function of the cohort, i.e.\ the expected fraction of the cohort surviving to age $x$, is then
\begin{equation} \label{eq:cohsurv}
   S(x) = \E\left[ S(x;Z) \right] = \E\left[\exp\left(-Z I(x) \right)\right] = L\left(I(x)\right),
\end{equation}
where $L$ denotes the Laplace transform of the frailty distribution (at birth),
\begin{equation}
   L(s)  = \E\left[\exp(-sZ)\right].
\end{equation}
Assuming sufficient regularity the (conditional) mean frailty of the cohort at age $x$ is given by
\begin{equation} \label{eq:meanfrailI}
   \E[Z|x]  = \frac{\E\left[Z S(x;Z) \right]}{S(x)} = \frac{\E\left[Z \exp\left(-Z I(x) \right) \right]}{L(I(x))}
   = - \frac{L'(I(x))}{L(I(x))} = \nu'(I(x)),
\end{equation}
where $'$ denotes differentiation and $\nu(s) = - \log L(s)$. It follows that the cohort intensity can be written as
\begin{equation}  \label{eq:cohint}
   \mu(x) =  - \frac{d}{dx}\log S(x) = \nu'(I(x)) I'(x) =  \E[Z|x] \mubase(x).
\end{equation}
The derivations above relating mean frailty and cohort intensity to the frailty distribution and baseline intensity
provide the usual tools for frailty modeling. For later use we make the additional
observation that mean frailty can also be expressed in terms
of the integrated cohort intensity,
\begin{equation}
   H(x) = \int_0^x \mu(u)du.
\end{equation}
The cohort survival function (\ref{eq:cohsurv}) provides the following link between $H$ and $I$,
\begin{equation}
   S(x) = \exp\left(-H(x)\right) = L\left(I(x)\right),
\end{equation}
from which we obtain $H(x)= \nu(I(x))$, and thereby $I(x) = \nu^{-1}(H(x))$. By (\ref{eq:meanfrailI}) and (\ref{eq:cohint}) we
then get the expression
\begin{equation}  \label{eq:cohintH}
   \mu(x) =  \nu'(\nu^{-1}\{H(x)\})\mubase(x).
\end{equation}
The idea later on is to replace $H$ by a non-parametric estimate obtained from data, i.e.\ the integrated
observed (empirical) death rates. The substitution disentangles the frailty distribution and the baseline intensity
which greatly simplifies the estimation procedure.

In order to apply the theory we need to identity suitable frailty distributions and their Laplace transforms.
Typically, the baseline intensity contains a scale parameter in which case it is customary to assume that the
frailty distribution has mean one to ensure identifiability. We will also make this assumption here.

\subsection{Gamma and inverse Gaussian frailty}  \label{sec:gaminvgau}
The frailty distributions most often employed in mortality modeling are the Gamma and inverse Gaussian distributions,
see e.g.\ \citet{vauetal79,hou84,manetal86,buthab04,jarkry11,spretal13}. Below we state their Laplace transforms and
comment on their use.

When frailty is Gamma-distributed with mean one and variance $\sigma^2$ the Laplace transform and mean frailty are given by
\begin{align}
   L(s) & =  \left( 1 + \sigma^2 s\right)^{-1/\sigma^2}, \\
 \E[Z|x] & = \left(1 + \sigma^2 I(x) \right)^{-1} = \exp\left(-\sigma^2 H(x)\right). \label{eq:EZgamma}
\end{align}
It is well-known that Gamma frailty in combination with Gompertz or Makeham
baseline intensity leads to a cohort intensity of logistic type, see e.g.\ Example 2.1 of \citet{jarkry11} for details.\footnote{A Gompertz intensity is of the form $\mubase(x)=\exp(a+bx)$, while
a Makeham intensity is of the form $\mubase(x)=\exp(a+bx) + c$. A Makeham intensity is also sometimes referred to as a Gompertz-Makeham intensity.}
This is also known as the Perks model, \citet{per32},  and it has been found to
describe old-age mortality very well, see e.g.\ \citet{thaetal98,tha99,caibladow06}.
Gamma-distributed frailty is also mathematically tractable and allows explicit calculations of many quantities of interest,
e.g.\ frailty among survivors at a given age is Gamma-distributed with known scale and shape parameters, cf.\ \citet{vauetal79}.

When frailty follows an inverse Gaussian distribution with mean one and variance $\sigma^2$ we have
for the Laplace transform and mean frailty
\begin{align}
   L(s) & = \exp\left[\frac{1-\sqrt{1+2\sigma^2 s}}{\sigma^2}\right], \label{eq:houLap}  \\
 \E[Z|x] & = \left(1 + 2\sigma^2 I(x) \right)^{-1/2} = \left( 1 + \sigma^2 H(x) \right)^{-1}. \label{eq:EZinvgau}
\end{align}
Similar to Gamma frailty inverse Gaussian frailty enjoys a number of nice theoretical properties, e.g.\ the
frailty distribution among survivors of a given age is again inverse Gaussian and an explicit
formula for the density exists, see \citet{hou84}. \citet{hou84} shows further that for inverse Gaussian frailty the coefficient
of variation among survivors is decreasing with age making the cohort more homogeneous as it gets older, while for Gamma frailty
the coefficient of variation is constant. Thus the impact of frailty is smaller for inverse Gaussian
frailty than for Gamma frailty. From (\ref{eq:EZgamma}) and (\ref{eq:EZinvgau}) we also see directly how
Gamma frailty "slows" down baseline intensity more than inverse Gaussian frailty.

It is generally found that Gamma frailty and the associated logistic form provides a better description of old-age mortality than inverse Gaussian frailty,
see e.g.\ \citet{buthab04,spretal13}. Furthermore, \citet{abbber07} show that for a large class of initial frailty distributions the frailty
distribution among survivors converges to a Gamma distribution as the integrated intensity tends to
infinity. Thus overall the Gamma distribution is a good default choice.

That being said, it is not immediately clear that this conclusion carries over to the case where frailty is introduced
to model increasing rates of improvement over time.

\subsection{Positive stable frailty} \label{sec:stablelaw}
\citet{hou86} introduced a family of generalized stable laws which include the Gamma and inverse Gaussian
distributions as special cases. The family is obtained by exponential tilting of stable densities with
index $\alpha \in [0,1)$. The stable laws themselves only have moments of order strictly less than $\alpha$, while moments
of all orders exist for the exponentially tilted densities. From the original three-parameter family we obtain a two-parameter family
by imposing the condition that mean frailty is one.

When $Z$ follows a generalized stable law with index $\alpha \in [0,1)$, mean one and variance $\sigma^2$ the Laplace
transform and mean frailty are given by\footnote{The stated formulaes are obtained from \citet{hou86} using
the parametrization $\theta = (1-\alpha)/\sigma^2$ and $\delta = [(1-\alpha)/\sigma^2]^{1-\alpha}$.}
\begin{align}
   L(s) & = \exp\left[\frac{1-\alpha}{\alpha}\left\{\frac{1-[1+\sigma^2 s/(1-\alpha)]^\alpha}{\sigma^2}\right\}\right], \\
 \E[Z|x] & = \left[1+\frac{\sigma^2}{1-\alpha}I(x)\right]^{\alpha-1} = \left[1 + \frac{\alpha}{1-\alpha}\sigma^2 H(x)\right]^{(\alpha-1)/\alpha}.
\end{align}
We note that for $\alpha = 0$ (defined by continuity) the generalized stable law specializes to the Gamma distribution, while for
$\alpha=1/2$ it specializes to the inverse Gaussian distribution. While Gamma and inverse Gaussian densities
are available in closed form, generally only series representations of (tilted) stable densities exist, cf.\ \citet{hou86} and Lemma XVII.6.1 of \citet{felII}.

In either case, we have simple closed form expressions for mean frailty in terms of integrated baseline and cohort intensity which is all we need
for estimation and forecasting purposes. Moreover, the fact that the parametric family spans the two most commonly used frailty distributions, Gamma and inverse Gaussian,
opens for a unified approach for joint estimation of frailty and baseline parameters.

The family has been further generalized by \citet{aal88,aal92} to include also negative $\alpha$. Negative values
of $\alpha$ correspond to compound Poisson distributions with positive probability of zero frailty.  This is important
for some applications when modeling the time to non-certain events, e.g.\ divorce or unemployment. The generalization
is of no use in our context where it implies immortality for some individuals. However, it could be useful for course-specific mortality modeling
as demonstrated by \citet{aal88}.

\section{Stochastic frailty models} \label{sec:SFM}
We first include frailty in a continuous-time model spanning multiple birth cohorts and show the effect on the dynamics of age-specific death rates over time.
Then we add frailty to stochastic mortality models and discuss estimation and forecasting of the resulting models.

Again, we start with the multiplicative frailty model. With a slight abuse of notation we will reuse the notation of Section~\ref{sec:frail}
with an added time parameter. Assume that the intensity for an individual of age $x$ at time $t$ with frailty $z$
has the form
\begin{equation}
   \mu(t,x;z) = z \mubase(t,x),
\end{equation}
where $\mubase$ is a baseline intensity describing the period (time) and age effect. The cohort intensity
is then given by
\begin{equation}
   \mu(t,x) = \E[Z|t,x] \mubase(t,x),
   \label{eq:pophazard2d}
\end{equation}
where $\E[Z|t,x]$ denotes the (conditional) mean frailty of the cohort of age $x$ at time $t$, i.e.\ the cohort
born at time $t-x$.

Assume that all cohorts have the same frailty distribution at birth, and denote the Laplace transform of this
common distribution by $L$. Let $\nu(s) = -\log L(s)$, and define the integrated baseline and cohort intensities as
\begin{align}
    I(t,x) & = \int_0^x \mubase(u+t-x,u)du,  \label{eq:Icoh} \\
    H(t,x) & = \int_0^x \mu(u+t-x,u)du.
\end{align}
That is, we follow a specific birth cohort over time and age. We have as before the relations
$H(t,x) = \nu\left(I(t,x)\right)$ and
\begin{equation}
   \E[Z|t,x]  = \nu'(I(t,x)) = \nu'\left(\nu^{-1}\left\{H(t,x)\right\}\right).  \label{eq:meanfrailcoh}
\end{equation}

Following the notation of \citet{bon05} we define the rate of
improvement in (senescent) mortality as $\rho(t,x) =  - \partial \log \mu(t,x) / \partial t$.
It follows from (\ref{eq:pophazard2d}) that
\begin{equation}
\rho(t,x)
 = -\frac{\partial \log \E[Z|t,x]}{\partial t}
  +  \rho_{\textrm{base}}(t,x),
   \label{eq:rhodef}
\end{equation}
where $\rho_{\textrm{base}}(t,x) =  - \partial \log \mubase(t,x) / \partial t$ denotes the rate of improvement of baseline intensity.

Suppose we model the period effect of mortality improvements by decreasing
age-specific baseline mortality, i.e.\ $\rho_{\textrm{base}} > 0$.
For fixed $x$, the integrated baseline intensity will then also be decreasing over time,
while the mean frailty will be {\em increasing} over time due to less and less selection.
From (\ref{eq:rhodef}) we see that this implies that the rate of improvement
of cohort mortality will be smaller than the rate of improvement at the individual level.

At old ages where death rates, and thereby selection, is high the change
in mean frailty over time can substantially offset improvements in baseline mortality causing
rates of improvement of cohort mortality to be close to zero. As
improvements continue to occur at the individual level the selection effect gradually disappears
and rates of improvement of cohort mortality get closer to baseline rates of improvement.
The resulting pattern of gradually changing rates of improvement of old-age mortality resembles what is seen in data.
This will be illustrated in the applications in Section~\ref{sec:app}.

\subsection{Data and terminology}
Data are assumed to be of the form of death counts, $D(t,x)$, and corresponding exposures, $E(t,x)$, for a range
of years, $t_{\min} \leq t \leq t_{\max}$, and ages, $x_{\min} \leq x \leq x_{\max}$. Data may or may not be gender-specific, but that will not be part of the notation.
$D(t,x)$ denotes the number of deaths occurring in calender year $t$ among people aged $[x,x+1)$. Correspondingly, $E(t,x)$
denotes the total number of years lived during calender year $t$ by people of age $[x,x+1)$.
For readers familiar with the Lexis diagram, $D(t,x)$ counts the number of deaths in the square $[t,t+1)\times [x,x+1)$ of the Lexis
diagram and $E(t,x)$ gives the corresponding exposure, i.e.\ we work with so-called \emph{A-groups}.

From the death counts and exposures we form the observed (empirical) death rates
\begin{equation}
    m(t,x) = \frac{D(t,x)}{E(t,x)}.
     \label{eq:death_rates}
\end{equation}
The death rate is an estimate of the cohort intensity, $\mu(t,x)$, which
for modeling purposes is assumed constant over the square $[t,t+1)\times [x,x+1)$ in the following.

\subsection{Fragilization of mortality models}
Consider as given a stochastic model for baseline mortality, $\mubase(t,x)$. For ease of presentation we will assume
that it is of the form
\begin{equation} \label{eq:basemodel}
   \mubase(t,x) = F(\theta_t,\eta_x),
\end{equation}
where $\theta_t$ and $\eta_x$ are, possibly multi-dimensional, time- and age-dependent quantities and $F$ describes the functional dependence. We could
have included also cohort effects and more general dependence structures if so desired.

The form (\ref{eq:basemodel}) is chosen with two applications mind: with $\theta_t = k_t$, $\eta_x = (a_x,b_x)$ and $F(k,(a,b)) = \exp(a+bk)$ we get the
Lee-Carter model; with $\eta_x=x$ we get parametric time-series models where the functional form of the age-profile is determined by $F$. In the parametric
case we could use for instance a Gompertz law, $F(\theta_t,x) = \exp(\theta^1_t + \theta^2_t x)$, or a logistic function,
$F(\theta_t,x) = \exp(\theta^1_t + \theta^2_t x)/(1+ \exp(\theta^1_t + \theta^2_t x))$. The latter is the model considered by \citet{caibladow06}
for aggregate mortality. As mentioned in Section~\ref{sec:gaminvgau} the logistic model already has a frailty interpretation,
and it is therefore not an obvious candidate for further fragilization; but it can nevertheless be done.

For presentational convenience we assume that both $\theta_t$ and $\eta_x$ need to be estimated from data, although as just seen $\eta_x$ may in fact be fixed.
Once estimated we can forecast $\theta_t$ by ARIMA time-series methods and thereby obtain forecasts of $\mubase$ by inserting the forecasted values of
$\theta_t$ together with the estimated values of $\eta_x$ into (\ref{eq:basemodel}).

In order to proceed we assume that we have available a procedure for maximum likelihood estimation of $(\theta_t,\eta_x)$ in the model
where death counts are independent with
\begin{equation}
      D(t,x) \sim \mbox{Poisson}\left( \mubase(t,x) E(t,x)\right).
\end{equation}
In the Lee-Carter case such a procedure is described in \citet{broetal02}.

With these building blocks in place we now consider the following "fragilized" version of the baseline model in which
death counts are independent with
\begin{align}
      D(t,x) & \sim \mbox{Poisson}\left( \mu(t,x) E(t,x)\right), \label{eq:frailPoisson1} \\
      \mu(t,x) & = \E[Z|t,x]\mubase(t,x),          \label{eq:frailPoisson2}
\end{align}
where $\mubase$ is given by (\ref{eq:basemodel}) and $\E[Z|t,x]$ denotes the conditional mean frailty of the cohort of age $x$ at time $t$.
It is assumed that the frailty distribution at birth is the same for all cohorts and that the distribution belongs to a family indexed by $\phi$.
The parameters of the model are thus $(\phi,\theta_t,\eta_x)$. The Laplace transform of the frailty distribution with index $\phi$ is denoted $L_\phi$, and
this is assumed available in explicit form. Further, as a matter of convention we assume that mean frailty is one at birth and we define $\nu_\phi = -\log L_\phi$.

Based on (\ref{eq:Icoh}) and (\ref{eq:meanfrailcoh}) we can write
\begin{align}
        \mu(t,x) & = \nu'_\phi(I(t,x))F(\theta_t,\eta_x),  \\
        I(t,x)   & = \sum_{u=0}^{x-1} F(\theta_{u+t-x},\eta_u),  \label{eq:IsumF}
\end{align}
and insert this into (\ref{eq:frailPoisson1}). In principle, we can estimate all parameters jointly from the resulting likelihood function.
However, the likelihood function is intractable with frailty and baseline parameters occurring in a complex mix. Consequently, estimation
has to be handled on a case-by-case basis depending on the choice of frailty and baseline model.
Below we propose an alternative, generally applicable pseudo-likelihood approach which greatly simplifies the estimation task.

\subsection{Pseudo-likelihood function} \label{sec:pseudo}
From (\ref{eq:meanfrailcoh}) we know that $\E[Z|t,x]=\nu'\left(\nu^{-1}\left\{H(t,x)\right\}\right)$, where $H$ is the integrated cohort intensity $\mu$. At first sight
this does not seem to help much since $H$ is at least as complicated as $I$. However, in contrast to $I$ we can obtain a (model free) estimate
of $H$ directly from data as
\begin{equation}
    \tilde{H}(t,x) = \sum_{u=0}^{x-1} m(u+t-x,u).  \label{eq:Htilde}
\end{equation}
We thus propose to base estimation of (\ref{eq:frailPoisson1})--(\ref{eq:frailPoisson2}) on a likelihood function in which the term $\E[Z|t,x]$ is replaced by
$\nu'(\nu^{-1}\{\tilde{H}(t,x)\})$. The resulting approximate likelihood function is referred to as the pseudo-likelihood function,\footnote{The idea of basing inference on a pseudo-likelihood function was first introduced in spatial statistics, see \citet{bes75}.}
\begin{align}
    L(\phi,\theta,\eta) & = \prod_{t,x}\frac{\lambda(t,x)^{D(t,x)}}{D(t,x)!}\exp(-\lambda(t,x)), \\
    \lambda(t,x) & = \nu'_\phi(\nu^{-1}_\phi\{\tilde{H}(t,x)\})F(\theta_t,\eta_x)E(t,x).
\end{align}
Formally, it corresponds to estimating the modified model
\begin{align}
      D(t,x) & \sim \mbox{Poisson}\left( \mu(t,x) E(t,x)\right), \label{eq:pseudo1} \\
      \mu(t,x) & = \nu'_\phi(\nu^{-1}_\phi\{\tilde{H}(t,x)\})F(\theta_t,\eta_x).         \label{eq:pseudo2}
\end{align}
In model (\ref{eq:pseudo1})--(\ref{eq:pseudo2}) $\mu$ is separable in frailty and baseline parameters and the model is therefore considerably easier to handle than (\ref{eq:frailPoisson1})--(\ref{eq:IsumF}).\footnote{Equation (\ref{eq:pseudo2}) might still look daunting, but is simplifies in specific cases. If frailty is assumed
Gamma distributed with variance $\sigma^2$ (and mean one) we get $\mu(t,x)  = \exp(-\sigma^2 \tilde{H}(t,x))F(\theta_t,\eta_x)$,
while inverse Gaussian frailty with variance $\sigma^2$ (and mean one) yields $\mu(t,x)  = F(\theta_t,\eta_x)/(1+ \sigma^2 \tilde{H}(t,x))$.}
The separability allows for simple joint estimation based on the estimation procedure for the baseline model.

\begin{figure}[h]
\begin{center}
\includegraphics[width=14cm]{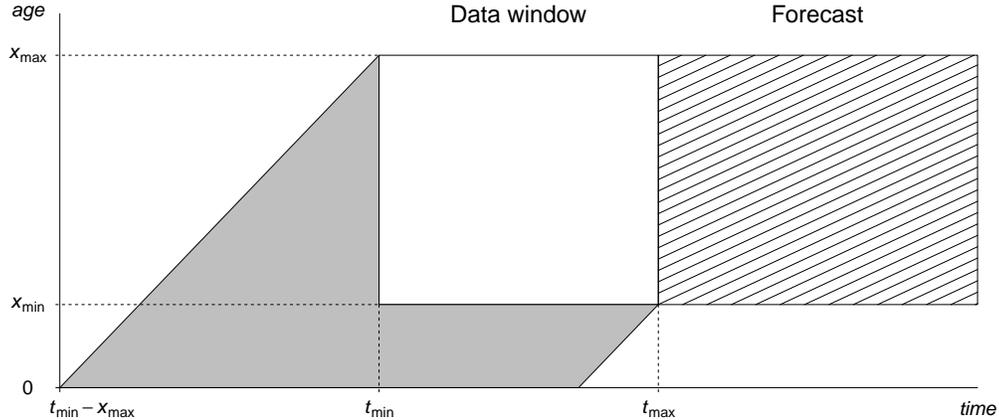}
\end{center}
\vspace*{-5mm}
\caption{Data are available for years between $t_{\min}$ and $t_{\max}$ and for ages between $x_{\min}$ and $x_{\max}$. The gray
area below and to the left of the data window illustrates the part of the trajectories needed for calculation of $\tilde{H}$ that
falls outside the data window. The shaded area to the right illustrates the years and ages for which we wish to forecast mortality intensities.}
\label{fig:datawindow}
\end{figure}

One issue remains before turning to estimation in Section~\ref{sec:MLest}. To form the (pseudo) likelihood function we need to compute
$\tilde{H}(t,x)$ for all $t$ and $x$ in the data window. However, these quantities depend in part
on data outside the data window. In principle, we need to know the death rates from birth
to the present/maximum age for all cohorts entering the estimation. The gray area of Figure~\ref{fig:datawindow} illustrates the "missing" death rates.

For the purpose of calculating $\tilde{H}$ we suggest to define the death rates before and below the data window by
\begin{equation}
   m(t,x) =
   \begin{cases}
   m(t_{\min},x) & \mbox{for } t < t_{\min} \mbox{ and } x_{\min} \leq x \leq x_{\max}, \\
   0             & \mbox{for } 0 \leq x < x_{\min}. \\
   \end{cases}
   \label{eq:mextension}
\end{equation}
This corresponds to saying that selection prior to $t_{\min}$ have happened according to initial rates (rather than actual rates),
and that all cohorts have mean frailty one at age $x_{\min}$ (rather than at birth). The latter assumption was also used by
Vaupel in the contributed discussion part of \citet{tha99}. With the extension of death rates $\tilde{H}$ and hence the likelihood function can be computed.

We note that in some situations, e.g.\ for many national data sets, we may have more data than we choose to use for estimation.
In these situations some or all of the "missing" death rates may in fact be available and hence could be used when calculating $\tilde{H}$.
In the present paper, however, we will calculate $\tilde{H}$ using the extension (\ref{eq:mextension}) even in this case. Assuming no selection
prior to $x_{\min}$ also makes forecasting more straightforward.

\subsection{Maximum likelihood estimation} \label{sec:MLest}
We here discuss various ways to obtain maximum likelihood estimates of model (\ref{eq:pseudo1})--(\ref{eq:pseudo2}) of Section~\ref{sec:pseudo}.
For computational efficiency and numeric stability one often considers the log-likelihood function (rather than the likelihood function),
\begin{equation}
    l(\phi,\theta,\eta) = \log L(\phi,\theta,\eta) = \sum_{t,x}\left\{D(t,x)\log \mu(t,x) - \mu(t,x)E(t,x)\right\} + \mbox{constant},  \label{eq:loglike}
\end{equation}
where $\mu$ is given by (\ref{eq:pseudo2}) and the constant depends on data only. The simplest and most general method to optimize (\ref{eq:loglike})
is to form the profile log-likelihood function
\begin{equation}
   l(\phi) =    l(\phi,\hat{\theta}(\phi),\hat{\eta}(\phi)),
\end{equation}
where $\hat{\theta}(\phi)$ and $\hat{\eta}(\phi)$ denote the maximum likelihood estimates of $\theta$ and $\eta$ for fixed frailty parameter $\phi$.
Assuming that we know how to perform maximum likelihood estimation of the baseline model, $\hat{\theta}(\phi)$ and $\hat{\eta}(\phi)$ can
be obtained by estimating the baseline model with death counts $D(t,x)$ and exposure $\nu'_\phi(\nu^{-1}_\phi\{\tilde{H}(t,x)\})E(t,x)$.
Since the frailty family is typically of low dimension, e.g.\ one or two dimensions, the profile log-likelihood function can normally be optimized reliably
by general purpose optimization routines, e.g.\ the {\tt optimize} or {\tt optim} routines in the free statistical software package $R$.

For a one-dimensional frailty family, e.g.\ Gamma with mean one and variance $\sigma^2$, we could use the following skeleton R-code
\begin{verbatim}
# model is a list with functions for calculating
# Htilde, mean frailty and MLE of the baseline model
frail.optim <- function(D,E,model) {

    Htilde <- model$Htilde(D,E)
    profile.loglike <- function(phi) {
        EZ  <- model$meanfrail(phi,Htilde)
        ret <- model$MLEbase(D,EZ*E)
        ret$value   # value of maximized log-likelihood
    }
    optret  <- optimize(profile.loglike,interval=c(0,2),maximum=TRUE)
    MLEphi  <- optret$maximum
    EZ      <- model$meanfrail(MLEphi,Htilde)
    ret     <- model$MLEbase(D,EZ*E)
    MLEbase <- ret$par  # MLE of baseline parameters
    list(phi=MLEphi,base=MLEbase)
}
\end{verbatim}

Alternatively, maximum likelihood estimation can also be implemented by a switching algorithm. Starting from an
initial value, $\phi_0$, of the frailty parameter and $i=1$, the algorithm consists in repeating
the following two steps until convergence. On each iteration $i$ is increased by one.
\begin{enumerate}
\item Calculate $(\theta_i,\eta_i)$ as the maximum likelihood
estimate with $\phi=\phi_{i-1}$ fixed.
\item Calculate $\phi_i$ as the maximum likelihood
estimate with $(\theta,\eta)=(\theta_i,\eta_i)$ fixed.
\end{enumerate}
The estimation in the first step is, as before, performed by estimating the baseline model with death
counts $D(t,x)$ and exposure $\nu'_{\phi_{i-1}}(\nu^{-1}_{\phi_{i-1}}\{\tilde{H}(t,x)\})E(t,x)$. The
estimation in the second step is performed by maximizing $l(\phi,\theta_i,\eta_i)$ as a function of $\phi$.
With the dimension of $\phi$ typically low, this is easy to do by either Newton-Raphson, steepest ascent, or,
in the one-dimensional case, even a simple bisection algorithm.

For the case of Gamma frailty with mean one and variance $\sigma^2$ we have for $l$ and its first two partial derivatives
with respect to $\sigma^2$
\begin{align*}
    l(\sigma^2,\theta,\eta) & = \sum_{t,x}\left\{-\sigma^2 D\tilde{H} - \exp(-\sigma^2 \tilde{H})F(\theta_t,\eta_x)E\right\} + \mbox{const.},  \\
    \frac{\partial l}{\partial \sigma^2}(\sigma^2,\theta,\eta) &=
    \sum_{t,x}\left\{-D\tilde{H} + \tilde{H}\exp(-\sigma^2 \tilde{H})F(\theta_t,\eta_x)E\right\},  \\
    \frac{\partial^2 l}{\partial (\sigma^2)^2}(\sigma^2,\theta,\eta) &=
     - \sum_{t,x}\tilde{H}^2\exp(-\sigma^2 \tilde{H})F(\theta_t,\eta_x)E < 0,
\end{align*}
where we have suppressed the dependence on $t$ and $x$ of $D$, $E$ and $\tilde{H}$ to simplify notation. The calculations
show that $l$ is strictly concave as a function of $\sigma^2$, and thereby that a local maximum is also a global maximum.
Hence any gradient method is guaranteed to converge to the global maximum. Note however that the maximum may
not be interior and therefore we do not necessarily have $\partial l/\partial \sigma^2=0$ at the maximum.

The switching algorithm will always converge to a (local) maximum. However, many iterations may be required in particular
if the likelihood has a "diagonal" ridge along frailty and baseline parameters, i.e.\ if frailty and baseline parameters to some
extent describe the same feature of data. The advantage of the switching algorithm is that it allows a more detailed analysis
which can guide the choice of optimization routine, like in the Gamma example above.

We finally note that estimation can also be carried out by Newton-Raphson sweeps over frailty and baseline parameters, similar
to the algorithm described in \citet{broetal02} for the Lee-Carter model. This may be more efficient in terms
of computing time, but it requires a substantial amount of tailor-made code for each combination of frailty and baseline model.

Overall, we find that optimization of the profile log-likelihood function is the method of choice. It is particularly well-suited
in the exploratory phase due to its flexibility and ease of implementation. In our experience it is also sufficiently
fast and robust to be of practical use.

\subsection{Forecasting}  \label{sec:fore}
We now assume that we have (maximum likelihood) estimates of $(\phi,\theta,\eta)$ and we denote these by $(\hat{\phi},\hat{\theta},\hat{\eta})$.
We will follow the usual approach in stochastic mortality modeling and forecast mortality based on a time-series model for $(\theta_t)$.
The time-series model is estimated on the basis of $(\hat{\theta}_t)_{t_{\min}\leq t \leq t_{\max}}$ treating these as observed, rather than estimated, quantities.
Typically a simple (multi-dimensional) random walk with drift is used, see e.g.\ \citet{leecar92} and \citet{caibladow06}, but models with more structure can also be used.

Assume that we have a forecast $(\bar{\theta}_t)_{t_{\max} < t \leq t_{\max}+h}$ for a given forecast horizon $h$.
The forecast can be either a deterministic (mean) forecast, or a stochastic realization from the time-series model.
The aim is to forecast baseline and cohort mortality for ages $x_{\min} \leq x \leq x_{\max}$; the forecast region
is illustrated as the shaded box to the right in Figure~\ref{fig:datawindow}.

We first note that the estimated $\mu$ can be written
\begin{equation}
    \hat{\mu}(t,x) = \nu'_{\hat{\phi}}(\nu^{-1}_{\hat{\phi}}\{\tilde{H}(t,x)\})F(\hat{\theta}_t,\hat{\eta}_x) =
      \nu'_{\hat{\phi}}(\tilde{I}(t,x))F(\hat{\theta}_t,\hat{\eta}_x),
\end{equation}
where $\tilde{I}(t,x) = \nu^{-1}_{\hat{\phi}}\{\tilde{H}(t,x)\}$. Expressing mean frailty in terms
of $\tilde{I}$ provides the necessary link for forecasting mean frailty, and thereby cohort mortality.

Baseline mortality is readily forecasted by inserting $\bar{\theta}$ and $\hat{\eta}$ into (\ref{eq:basemodel}),
\begin{equation}
   \mubase(t,x) = F(\bar{\theta}_t,\hat{\eta}_x),
\end{equation}
while cohort mortality is forecasted by
\begin{equation}
   \mu(t,x) = \nu'_{\hat{\phi}}(\tilde{I}(t,x))F(\bar{\theta}_t,\hat{\eta}_x),
\end{equation}
where $\tilde{I}$ in the forecast region is given by the recursion
\begin{equation}
   \tilde{I}(t,x) =
   \begin{cases}
   0             & \mbox{for } x = x_{\min}, \\
   \tilde{I}(t_{\max},x-1)+F(\hat{\theta}_{t_{\max}},\hat{\eta}_{x-1})      & \mbox{for } x > x_{\min}, t=t_{\max}+1, \\
   \tilde{I}(t-1,x-1)+F(\bar{\theta}_{t-1},\hat{\eta}_{x-1})      & \mbox{for } x > x_{\min}, t > t_{\max}+1. \\
   \end{cases}
   \label{eq:Itildefor}
\end{equation}
Note that in the data window $\tilde{I}$ is defined by transformation of $\tilde{H}$ to ensure consistency
with the estimated model, while in the forecast region it is defined recursively in terms of the forecasted
baseline mortality.

For Gamma frailty with mean one and estimated variance $\hat{\sigma}^2$ we will forecast cohort mortality by
\begin{equation}
       \mu(t,x) = \frac {F(\bar{\theta}_t,\hat{\eta}_x)}{1 + \hat{\sigma}^2 \tilde{I}(t,x)},
\end{equation}
where $\tilde{I}$ in the data window is given by $\tilde{I} = [\exp(\hat{\sigma}^2 H)-1]/\hat{\sigma}^2$. Inverse Gaussian frailty
with mean one and estimated variance $\hat{\sigma}^2$ yields the cohort mortality forecast
\begin{equation}
       \mu(t,x) = \frac {F(\bar{\theta}_t,\hat{\eta}_x)}{\sqrt{1 + 2\hat{\sigma}^2 \tilde{I}(t,x)}},
\end{equation}
where $\tilde{I}$ in the data window is given by $\tilde{I} = [(1+\hat{\sigma}^2H)^2-1]/2\hat{\sigma}^2$.

\section{Applications to US male mortality}  \label{sec:app}
Broadly speaking stochastic mortality models fall in two main categories:
parametric time-series models and semi-parametric models of Lee-Carter type. We consider two applications,
one for each of these model classes. The applications are illustrative examples
highlighting different aspects of the theory rather than full-blown statistical analyses.

In the first application we take a standard Makeham mortality intensity and show how the fit can be
improved for different age groups by adding frailty. To give a flavor of the possibilities
we consider models of the generalized, additive form presented in Section~\ref{sec:genSFM}
in combination with the two-dimensional family of stable laws of Section~\ref{sec:stablelaw}.
In the second application we add Gamma frailty to the Lee-Carter model and illustrate the
effect on the forecasts.

Both applications use US male mortality data from 1950--2010 available at the Human Mortality Database (www.mortality.org).
The first application compares a forecast based on the initial half of data with the actual mortality experience.
The second application uses the latter half of data and makes a forecast into the future.

\subsection{Parametric time-series application} \label{sec:app1}
\begin{figure}[h]
\begin{center}
\includegraphics[height=7.5cm]{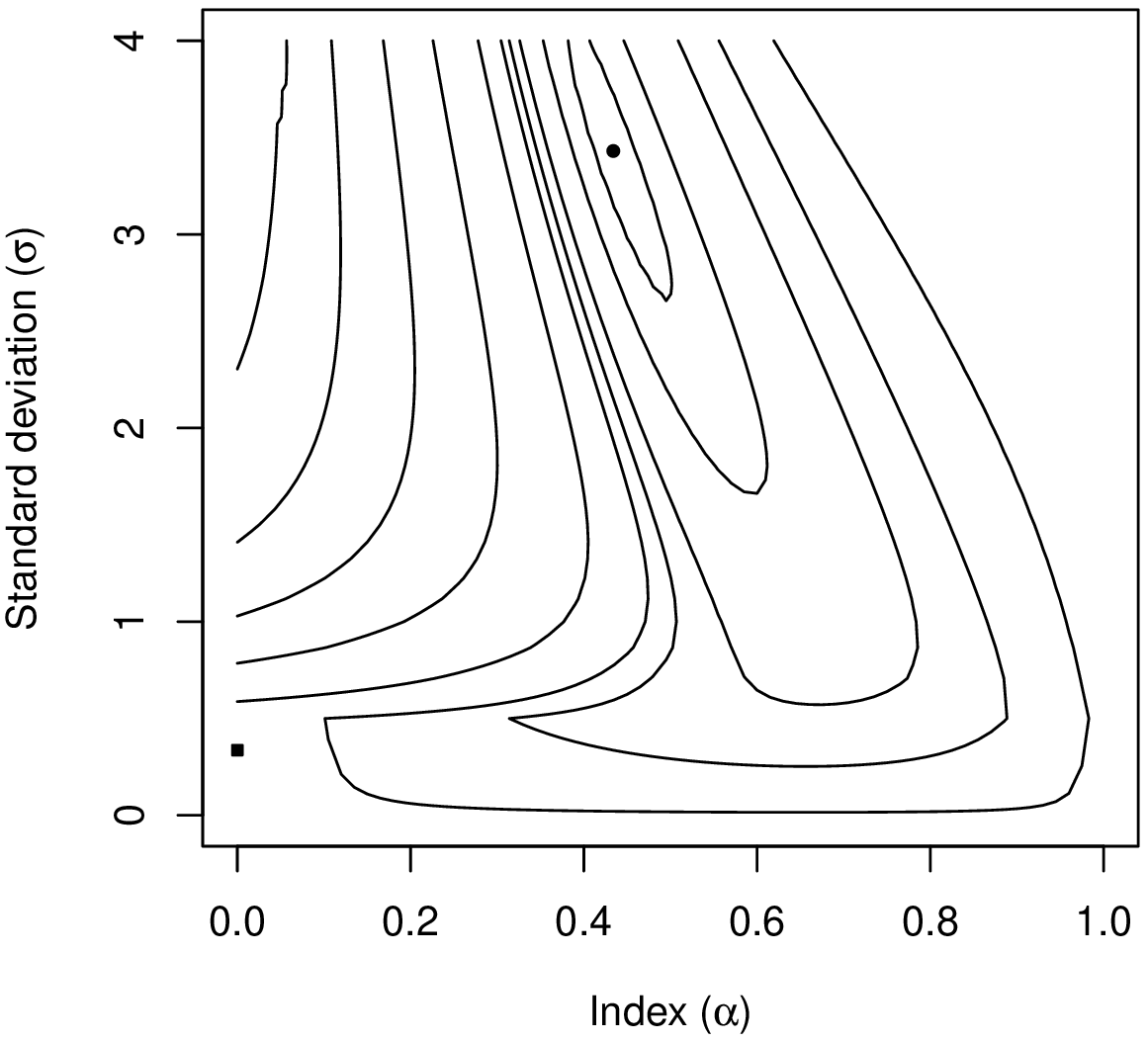}
\hfill
\includegraphics[height=7.5cm]{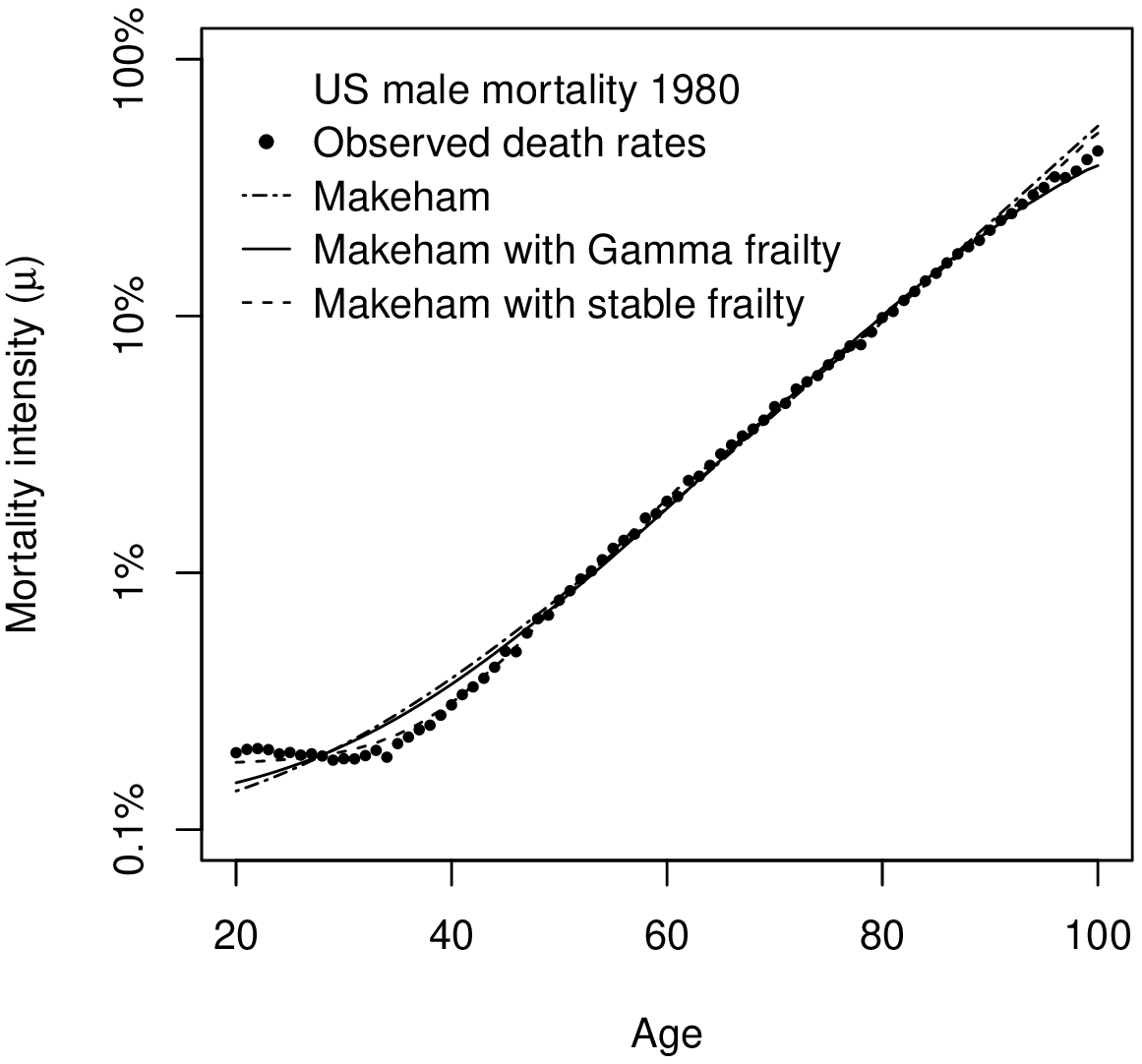}
\end{center}
\vspace*{-5mm}
\caption{Left panel shows a contour plot of the profile log-likelihood function, $l(\alpha,\sigma)$, of model (\ref{eq:sfmtsapp1})--(\ref{eq:sfmtsapp2}).
The dot marks the maximal value, $(\hat{\alpha},\hat{\sigma})$, and the box marks the maximal value along
the axis $\alpha=0$ (Gamma frailty), $\hat{\sigma}_{\Gamma}$. Right panel shows observed death rates in 1980 for US males aged 20 to 100 with a Makeham fit superimposed. Also shown
are the fits of model (\ref{eq:sfmtsapp1})--(\ref{eq:sfmtsapp2}) with optimal Gamma and stable frailty, respectively.}
\label{fig:TScontour}
\end{figure}

The classical Makeham, or Gompertz-Makeham, mortality law states that age-specific mortality
can be described by the functional form
\begin{equation}
  \mu(x) = \exp(\theta^1 + \theta^2 x ) + \exp(\zeta),
\end{equation}
where $x$ denotes age. The additive structure suggests an interpretation of mortality
as composed of an age-dependent (senescent) part describing mortality as people grow older,
and a part describing background mortality, e.g.\ accidents, which happen independent of age.
Despite its simplicity the Makeham intensity provides a reasonable description for a wide
range of adult ages. However, for the very old and also for younger adults
the description is not adequate. The right panel of Figure~\ref{fig:TScontour} shows
observed death rates in 1980 for US males aged 20 to 100 with a fitted Makeham curve imposed
(and two other curves which will be explained below). It is seen that the Makeham curve overstates
old-age mortality and that the curvature is wrong for the younger ages.

For old-age mortality, the lack of fit of the Makeham law is well-known
and logistic-type models have been proposed instead, see e.g.\ \citet{thaetal98}.
In \citet{caibladow06} a logit-model for death probabilities (rather than rates)
is suggested and applied to England \& Wales males above age 60 for which it fits well.
However, in \citet{caietal09} potential problems of fitting US male data
with the pure logit-model is reported and a variant is considered.

\subsubsection{Stochastic frailty model}
As an alternative to the logit-model and its variants we here consider
the stochastic frailty model
\begin{align}
      D(t,x)       & \sim \mbox{Poisson}\left( \mu(t,x) E(t,x)\right),  \label{eq:sfmtsapp1}\\
      \mu(t,x)     & = \E[Z|t,x]\exp(\theta_t^1 + \theta_t^2 x ) + \exp(\zeta_t),   \label{eq:sfmtsapp2}
\end{align}
where the distribution of $Z$ belongs to the positive stable family with index $\alpha\in [0,1)$ and variance $\sigma^2$,
described in Section~\ref{sec:stablelaw}. The model is of the additive
form introduced in Section~\ref{sec:genSFM} with frailty acting only on the senescent part of mortality.
With this structure we keep the interpretation of the second term as background mortality to which everyone is equally susceptible.

We recall from Section~\ref{sec:stablelaw} that the stable family includes both Gamma ($\alpha=0$) and
inverse Gaussian ($\alpha=1/2$) as special cases. Gamma frailty leads to a logistic-type model, akin to the one used in \citet{tha99},
while values of $\alpha$ greater than zero yield old-age mortality "in between" logistic and exponential.
The Makeham model with time-varying parameters is also included and corresponds to $\sigma^2=0$ (for all values of $\alpha$).

We apply the model to US male mortality data for the period 1950--1980 and ages 20--100.
The left panel of Figure~\ref{fig:TScontour} shows a contour plot of the profile log-likelihood function,
\begin{equation}
  l(\alpha,\sigma) = \log L(\alpha,\sigma,\hat{\theta}(\alpha,\sigma),\hat{\eta}(\alpha,\sigma)),
\end{equation}
where $L$ is the pseudo-likelihood function of Section~\ref{sec:gpseudo} and $\hat{\theta}(\alpha,\sigma)$ and $\hat{\eta}(\alpha,\sigma)$
denote the maximum likelihood estimates for fixed value of frailty parameters. The profile log-likelihood function is
calculated by the EM-algorithm of Section~\ref{sec:gMLE}.

Interestingly, the profile log-likelihood is maximized for $(\hat{\alpha},\hat{\sigma}^2) = (0.434,11.770)$. This implies
that the best fit is obtained for a model which is quite far from a logistic form. A closer look at data reveals why
this is the case. The right panel of Figure~\ref{fig:TScontour} shows the observed death rates in 1980 with three fitted curves
imposed. The curve labeled 'Makeham with stable frailty' is the one corresponding to the MLE, while
the curve labeled 'Makeham' is the fitted Makeham curve mentioned earlier. If we
restrict attention to Gamma frailty, i.e.\ maximize the profile log-likelihood along the axis $\alpha=0$, we find
the maximum $\hat{\sigma}_{\Gamma}^2=0.113$. The corresponding fitted curve is the one labeled 'Makeham with Gamma frailty'.

Compared to the Makeham curve, the MLE curve gives an improved fit at young ages and a similar fit at old ages,
while the Gamma curve gives an improved fit at old ages and a similar fit at young ages. Since
the exposure is much bigger at young ages than at old ages the global MLE is the one fitting young ages the
best. Ideally we would like a model which combines the MLE fit for young ages with the Gamma fit at old ages.
Note that the observed death rates are actually decreasing from age 20 to around age 30. This however
cannot be captured by any of the models considered.

\subsubsection{Forecast}
\begin{figure}[h]
\begin{center}
\includegraphics[height=7.5cm]{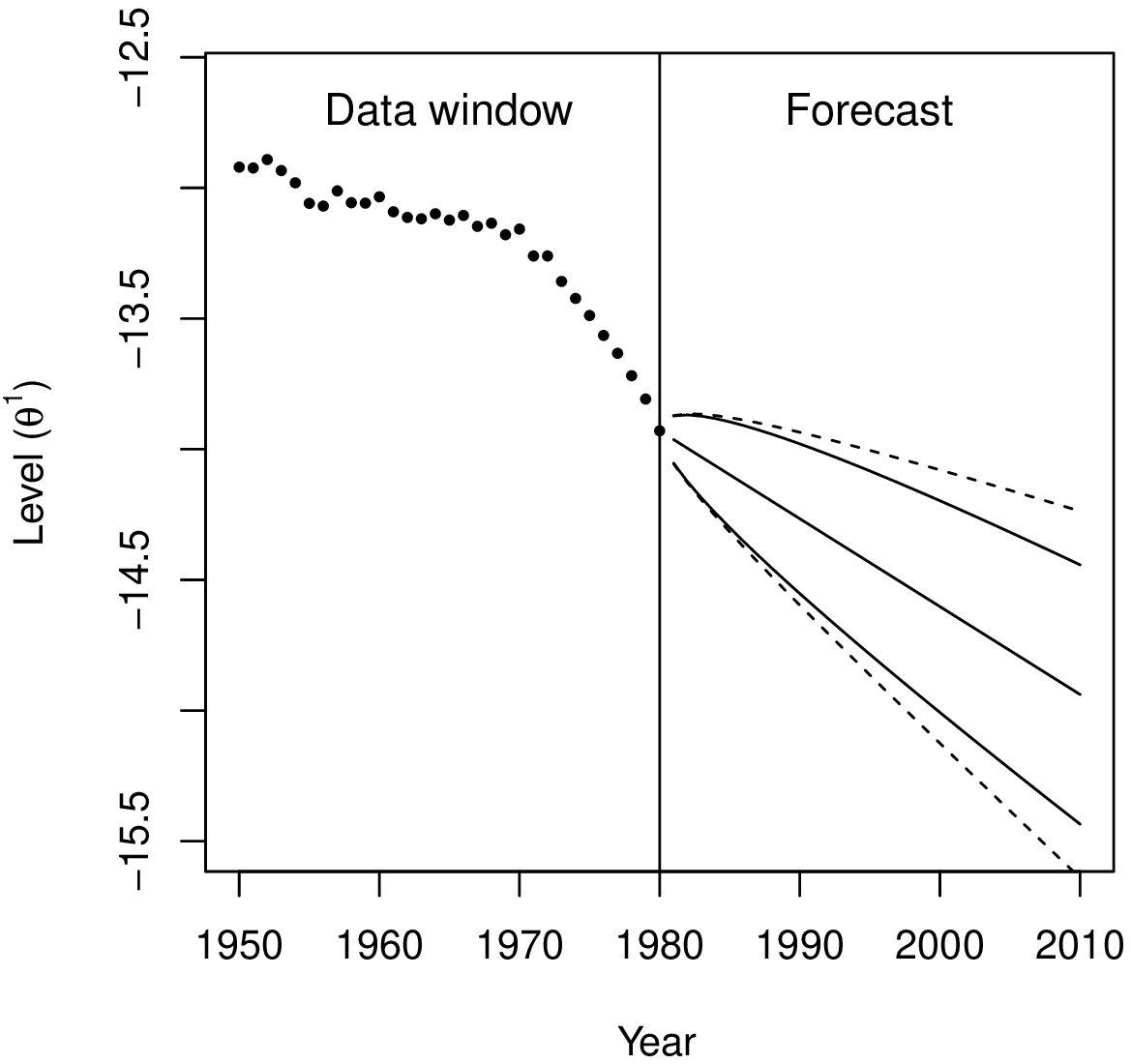}
\hfill
\includegraphics[height=7.5cm]{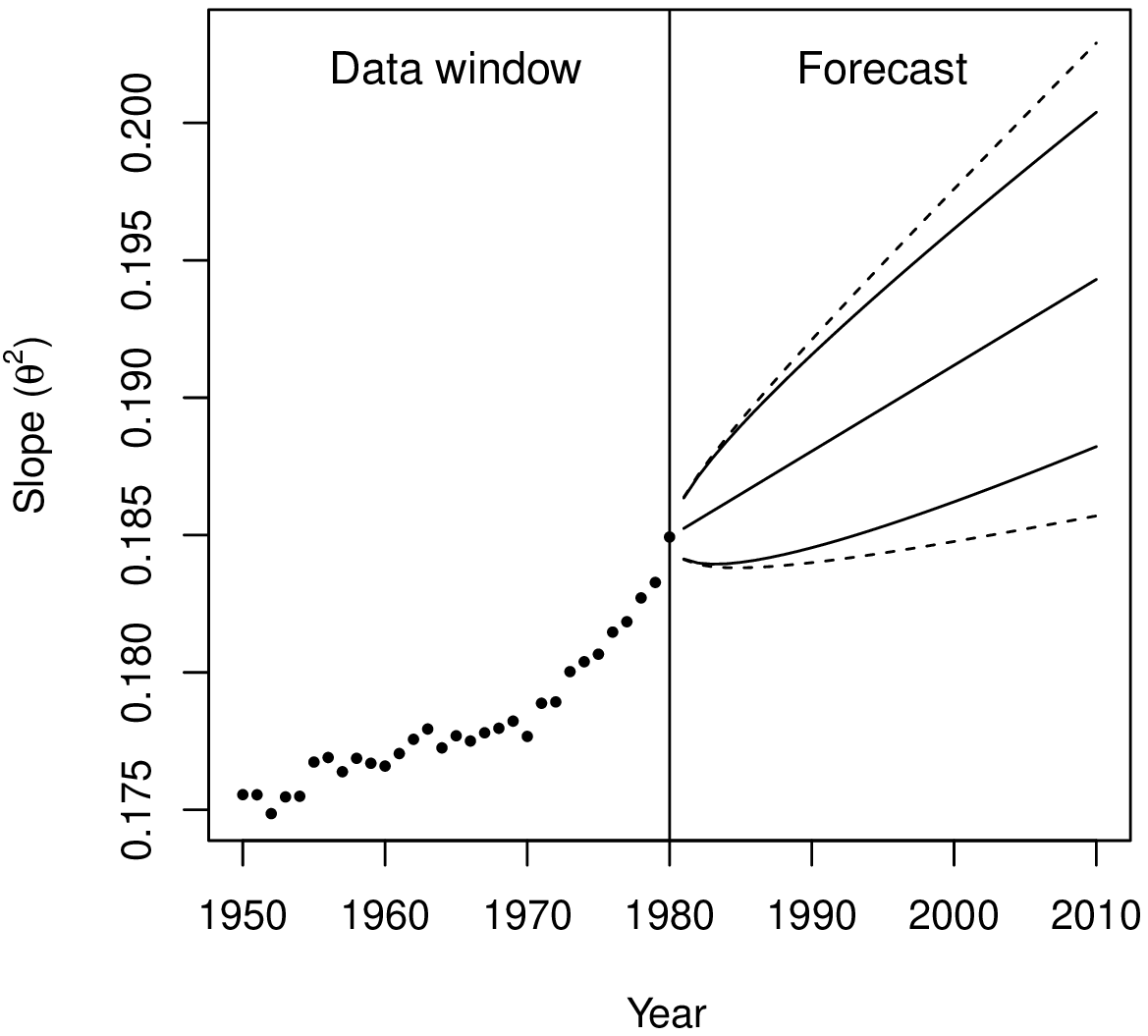}
\end{center}
\vspace*{-5mm}
\caption{The dots show maximum likelihood estimates of level, $\theta^1$, and slope, $\theta^2$, parameters of model (\ref{eq:sfmtsapp1})--(\ref{eq:sfmtsapp2}) with stable frailty.
Mean forecasts based on model (\ref{eq:RWtheta}) are shown as solid lines. 95\%-confidence intervals with and without
parameter uncertainty on $\xi$ (drift) are shown as dashed and solid lines respectively.}
\label{fig:TSparest}
\end{figure}

Figure~\ref{fig:TSparest} shows the estimated and forecasted level, $\theta^1$, and slope, $\theta^2$, parameters of individual senescent mortality.
The forecast is produced as the mean forecast of a fitted two-dimensional random walk with drift,
\begin{equation}
       \theta_t = \theta_{t-1} + \xi + U_t, \label{eq:RWtheta}
\end{equation}
where $\xi$ is the drift and $U_t$  are independent, identically distributed two-dimensional normal variables with mean zero.
Qualitatively similar plots hold for the Gamma frailty model (not shown).

It appears that there is a structural break around 1970 where the annual drift in level and slope changes. With hindsight we now know
that this change was genuine and that using only the period 1970--1980 would have resulted in a better forecast. Imagine standing
in 1980, however, it is less obvious that we would have used only the last 10 years of data when forecasting. To illustrate the forecast
the method would likely have produced in 1980 we therefore use the mean forecast based on the average drift for the full period 1950--1980, as shown in Figure~\ref{fig:TSparest}.

Cohort mortality with stable frailty is forecasted by
\begin{equation}
  \mu(t,x) = \exp(\bar{\theta}_t^1 + \bar{\theta}_t^2 x )\left[1+\frac{11.770}{1-0.434}\tilde{I}(t,x)\right]^{0.434-1} + \exp(\bar{\zeta}_t),
\end{equation}
where $\bar{\theta}$ denotes the mean forecast of $\theta$, cf.\ Section~\ref{sec:gforecast}. Background mortality shows
very little variation over the data window and is therefore kept fixed at the last value in the forecast, $\bar{\zeta_t} = \hat{\zeta}_{1980}$.
Similarly, cohort mortality with Gamma frailty is forecasted by
\begin{equation}
  \mu(t,x) = \frac {\exp(\bar{\theta}_t^1 + \bar{\theta}_t^2 x )}{1 + 0.113 \tilde{I}(t,x)} + \exp(\bar{\zeta}_t).
\end{equation}

\begin{figure}[h]
\begin{center}
\includegraphics[height=7.5cm]{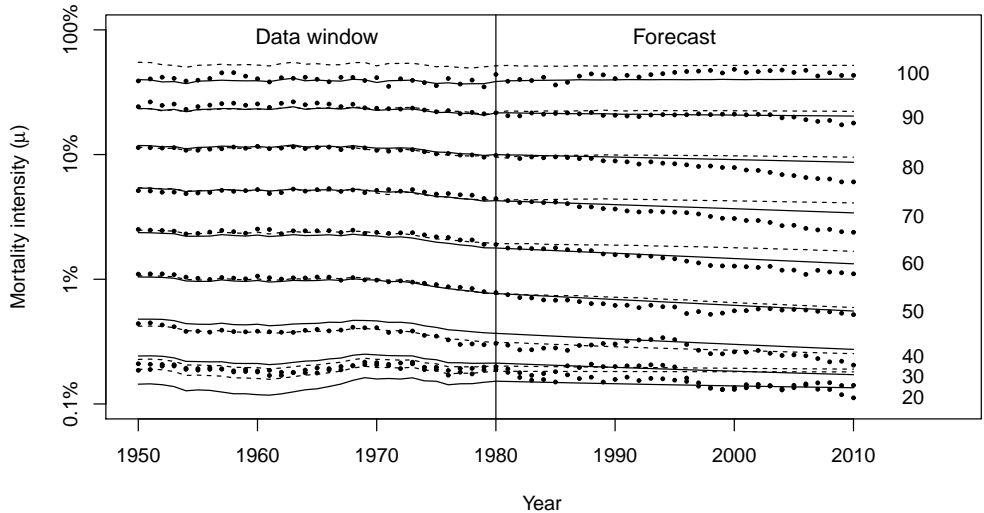}
\end{center}
\vspace*{-5mm}
\caption{The dots show observed death rates for US males from 1950 to 2010 for ages 20, 30, $\ldots$, 100 years.
The fit and forecast of model (\ref{eq:sfmtsapp1})--(\ref{eq:sfmtsapp2}) with optimal Gamma and stable frailty are shown
as solid and dashed lines respectively}
\label{fig:TSfitfor}
\end{figure}

The fit and forecast with the two frailty specifications are shown in Figure~\ref{fig:TSfitfor} together with
observed death rates from 1950 to 2010. In the data window, the model with stable frailty is seen to provide a very good fit
for all ages except the oldest (age 100), while Gamma frailty leads to a good fit for all ages above 50, but
a poor fit below. None of the models are able to fully predict the improvements that occurred
from 1980 onwards in particular for ages 60--80. Of the two models the Gamma frailty model does in fact come closest due to its
more log-concave forecasts, cf.\ the discussion of Gamma versus inverse Gaussian frailty in Section~\ref{sec:gaminvgau}.

Overall, the example shows that adding frailty can substantially improve the fit of a simple baseline model.
However, as with all parsimonious models the fit is not perfect, in particular when we consider an age span as wide as 20 to 100 years.
The example also shows that regarding both fit and forecast of old-age mortality Gamma frailty performs better than (other)
stable frailties.

\subsection{Lee-Carter application} \label{sec:app2}
The model of \citet{leecar92} assumes a log-bilinear structure of mortality,
\begin{equation}
     \log \mu(t,x) = a_x + b_x k_t,   \label{eq:app2LC}
\end{equation}
where $a$ and $b$ are age-specific parameters and $k$ is a time-varying index.
The index is typically modeled as a random walk with drift,
\begin{equation}
   k_t = k_{t-1} + \xi + U_t,      \label{eq:app2RW}
\end{equation}
where $\xi$ is the drift and $U_t$ are independent, identically distributed
normal variables with mean zero. The combination of (\ref{eq:app2LC}) and (\ref{eq:app2RW}) implies that
forecasted age-specific death rates decay exponentially at a constant rate. However, the experience
of old-age mortality has shown increasing rates of improvement in many countries and, consequently,
forecasts based on the Lee-Carter methodology have a tendency to underestimate the actual gains.
In the example to follow we illustrate how the addition of frailty can be used to improve
the forecast of old-age mortality.

We consider the Poisson version of the Lee-Carter model,
\begin{equation}
  D(t,x) \sim \mbox{Poisson}(\exp( a_x + b_x k_t)E(t,x)),
\end{equation}
and we use the algorithm of \citet{broetal02} to obtain the maximum likelihood estimates of the parameters
under the usual identifiability constraints,
\begin{equation}
    \sum_t k_t = 0 \quad \mbox{and} \quad \sum_x b_x = 1.
\end{equation}

\begin{figure}[h]
\begin{center}
\includegraphics[height=7.5cm]{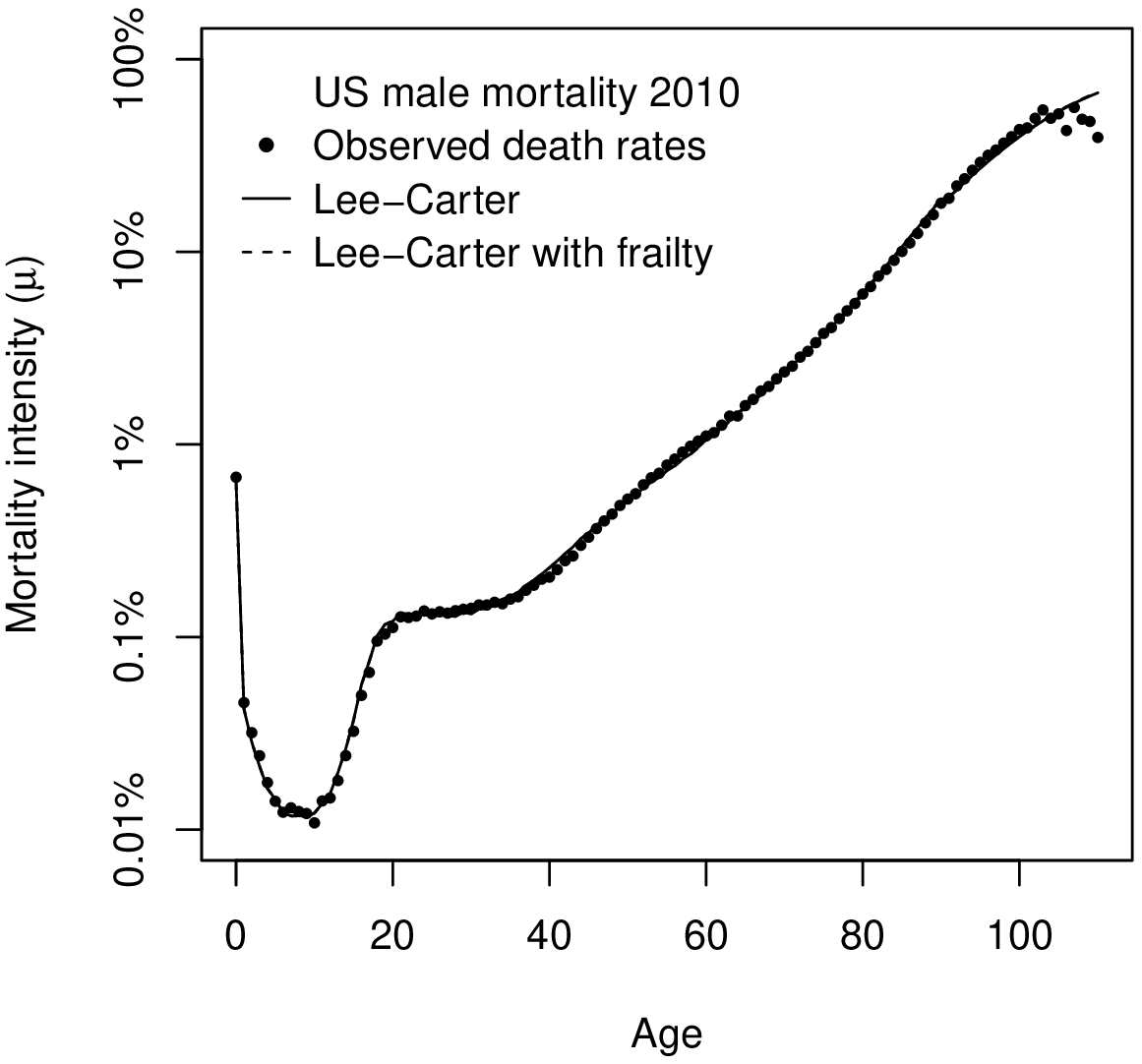}
\hfill
\includegraphics[height=7.5cm]{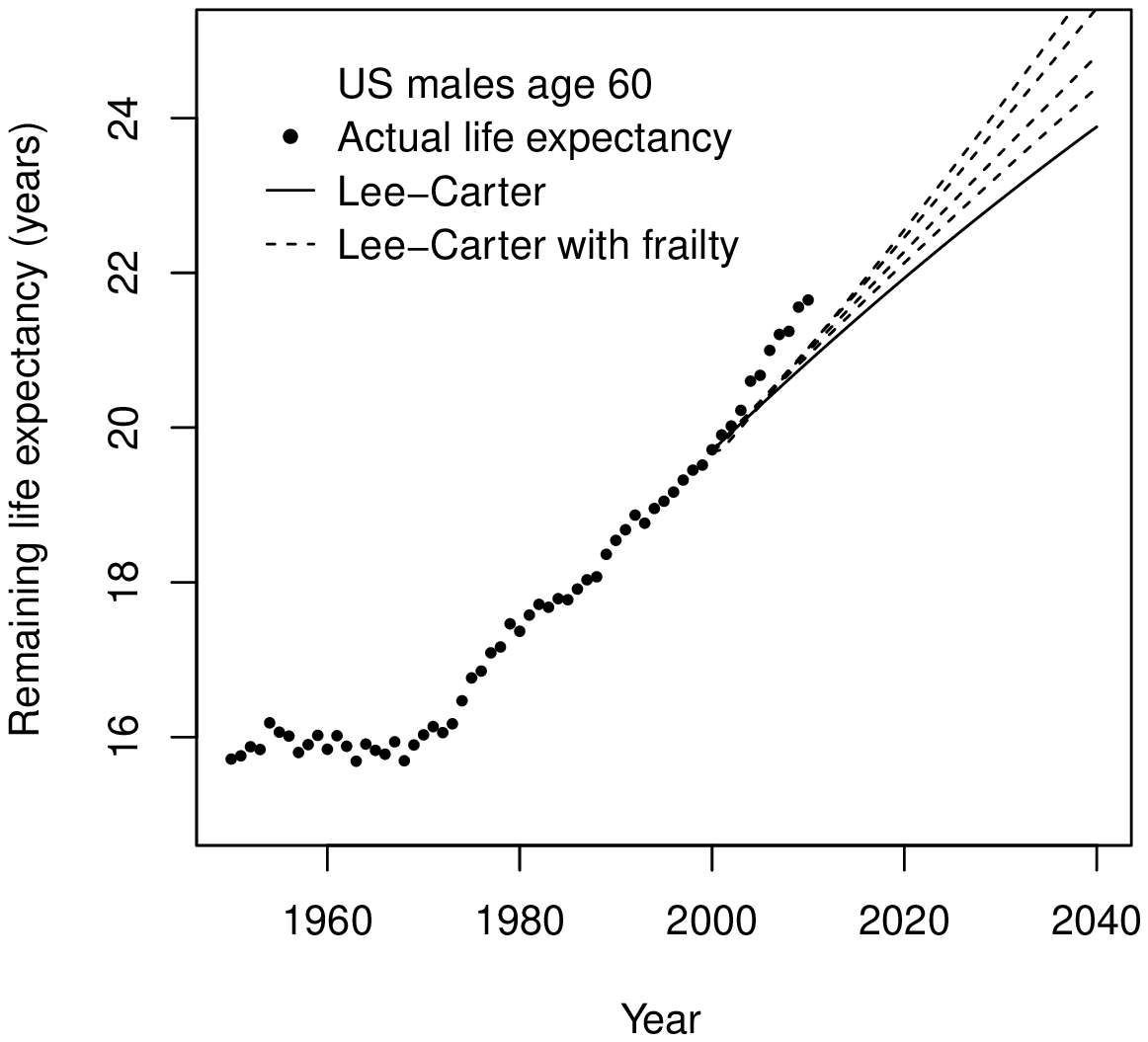}
\end{center}
\vspace*{-5mm}
\caption{Left panel shows observed death rates in 2010 for US males ages 0 to 110. The fitted value for model (\ref{eq:sfmLCapp1})--(\ref{eq:sfmLCapp2}) without frailty (solid) and with $\hat{\sigma}^2=0.73$ (dashed) are superimposed. Logistic extension above age 90. Right panel shows actual and forecasted life expectancy for US males age 60. Forecasts are based
on model (\ref{eq:sfmLCapp1})--(\ref{eq:sfmLCapp2}) without (solid) and with (dashed) frailty and estimation period 1970--2000. The frailty forecasts
are for $\sigma^2$ = 0.25, 0.5, 1.0 and 1.5 with higher values of $\sigma^2$ corresponding to higher forecasts.}
\label{fig:LCfit}
\end{figure}

We will use the Lee-Carter model with and without frailty to model mortality for ages 0 to 90, and assume a logistic
form for higher ages,\footnote{For each $t$, $c_t$ and $d_t$ are estimated by least squares based on the relation $\logit\mu(t,x) = c_t + d_t x  + \mbox{error}$,
for ages $x=70,71,\ldots,90$.}
\begin{equation}
    \mu(t,x) = \frac{\exp(c_t + d_t x)}{1+ \exp(c_t + d_t x)} \quad \mbox{for } x=91,\ldots,110.
\end{equation}
The extrapolation at the oldest ages is necessary to obtain reliable and stable rates, both historically where data are sparse and
in forecasts. The left panel of Figure~\ref{fig:LCfit} shows the fit in 2010 for the full age-span from 0 to 110 years.
The plot illustrates the flexibility and better fit to data of semi-parametric models, like the Lee-Carter model,
over parametric models like the ones considered in Section~\ref{sec:app1}. Parametric models, on the other hand,
generally produce forecasts which better preserve the overall structure of data.

The four Lee-Carter forecasts used in the introductory example in Section~\ref{sec:intro} are produced as described above. The mean of the index, $\bar{k}_t$, is used for forecasting
assuming an underlying random walk with drift. The period life expectancy used for illustration
is calculated by the following formula with $x_0=60$
\begin{equation}
   e_{x_0}(t) = \int_{x_0}^{110} \exp\left( - \int_{x_0}^x \mu(t,\lfloor y \rfloor)dy  \right)dx,
\end{equation}
where  $\lfloor y \rfloor$ denotes the integer part of $y$. Arguably, the cohort life expectancy
taking future improvements into account is of more interested in practise. However, for our purposes the
period life expectancy is more useful since it can be compared with the actual experience.

\subsubsection{Stochastic frailty model}
Consider the following multiplicative stochastic frailty model
\begin{align}
      D(t,x)       & \sim \mbox{Poisson}\left( \mu(t,x) E(t,x)\right),  \label{eq:sfmLCapp1}\\
      \mu(t,x)     & = \E[Z|t,x]\exp( a_x + b_x k_t),   \label{eq:sfmLCapp2}
\end{align}
where $Z$ is assumed to be Gamma distributed with mean one and variance $\sigma^2$.
Due to the assumption of Gamma frailty we have the relations
\begin{equation}
   \E[Z|t,x] = \left(1 + \sigma^2 I(t,x) \right)^{-1} = \exp\left(-\sigma^2 H(t,x)\right), \label{eq:EZgammaLCapp}
\end{equation}
where $H$ and $I$ denote integrated cohort and baseline mortality respectively. For $\sigma^2=0$
we have the ordinary Lee-Carter model.

We will deviate slightly from the theory presented so far and use the period version of $H$ and $I$
rather than the cohort version previously used. Hence we will replace $H$ by
the quantity
\begin{equation}
    \check{H}(t,x) = \sum_{u=0}^{x-1} m(t,u),  \label{eq:Hcheck}
\end{equation}
when estimating the model. Similarly, we calculate and forecast $I$ by its period version $\check{I}$.
Note that with the period version we do not need to extend the data window.

In the context of the Lee-Carter model the period version appears to stabilize the estimates and forecasts,
and this is why we use it. We also note that the use of Gamma frailty on a period basis is consistent with the logistic extension
above age 90.

For fixed value of $\sigma^2$ model (\ref{eq:sfmLCapp1})--(\ref{eq:sfmLCapp2}) is estimated as an ordinary
Lee-Carter model with exposure $\exp(-\sigma^2 \check{H}(t,x))E(t,x)$. With the large number
of parameters of the Lee-Carter model the addition of $\sigma^2$ makes little difference for the fit.
This is illustrated in the left panel of Figure~\ref{fig:LCfit} where the fit of the Lee-Carter model
with and without frailty is indistinguishable.

Since frailty is introduced with the specific aim of improving the forecast
we propose to estimate $\sigma^2$ with a back-test approach, rather than by maximum likelihood estimation.
Specifically, we define the forecast fit by
\begin{equation}
    f(\sigma^2) = \sum_{t=2001}^{2010} \sum_{x=0}^{90} \left\{D(t,x)\log \mu(t,x;\sigma^2) - \mu(t,x;\sigma^2)E(t,x)\right\},  \label{eq:loglikeLCapp}
\end{equation}
where $\mu(t,x;\sigma^2)$ denotes the forecast of model (\ref{eq:sfmLCapp1})--(\ref{eq:sfmLCapp2}) estimated with
data from 1970--2000 and ages 0--90 for fixed value of $\sigma^2$. The model is forecasted as described in Section~\ref{sec:forLCapp} below.
Plotting $f$ shows a unimodal function which is maximized at $\hat{\sigma}^2=0.73$ (plot not shown).
The chosen approach is simple and serves as illustration, but many other choices for measure of fit and data period could have been made.

The right panel of Figure~\ref{fig:LCfit} shows the life expectancy of US males age 60 resulting from
the forecast $\mu(t,x;\sigma^2)$ for different values of $\sigma^2$.
The impact of frailty is initially modest but it gradually leads to substantial differences. Yet, even the presence
of frailty cannot fully explain the rapid increase in life expectancy of the last decade. Note that the optimal
frailty variance is not the one leading to the highest forecast, but the one giving the best fit overall for ages 0--90 in
the ten years, 2001--2010, following the estimation period.

\subsubsection{Forecast}  \label{sec:forLCapp}
Forecasts of model (\ref{eq:sfmLCapp1})--(\ref{eq:sfmLCapp2}) are obtained by
\begin{equation}
   \mu(t,x) = \frac{\exp( \hat{a}_x + \hat{b}_x \bar{k}_t)}{1+\hat{\sigma}^2\check{I}(t,x)},
\end{equation}
where $\hat{a}_x$, $\hat{b}_x$ and $\hat{\sigma}^2$ denote estimated parameters and $\bar{k}_t$ is
the forecasted index, either deterministic or stochastic. The integrated baseline intensity
is given by $\check{I}(t,x) = [\exp(\hat{\sigma}^2 \check{H}(t,x)) -1]/\hat{\sigma}^2$ in the data window
and by
\begin{equation}
   \check{I}(t,x) = \sum_{u=0}^{x-1}\exp( \hat{a}_u + \hat{b}_u \bar{k}_t)
\end{equation}
in the forecast.

\begin{figure}[h]
\begin{center}
\includegraphics[height=7.5cm]{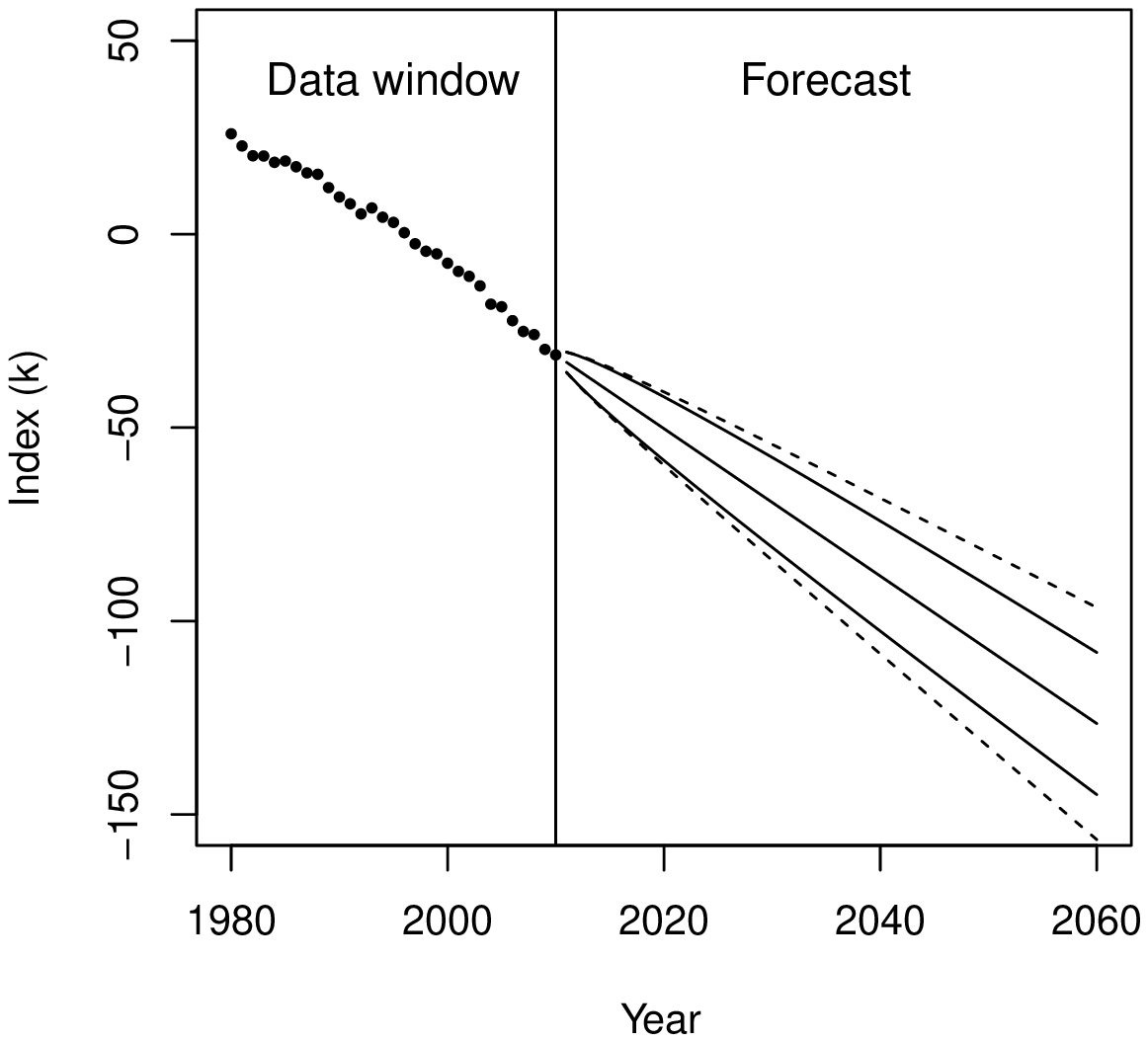}
\hfill
\includegraphics[height=7.5cm]{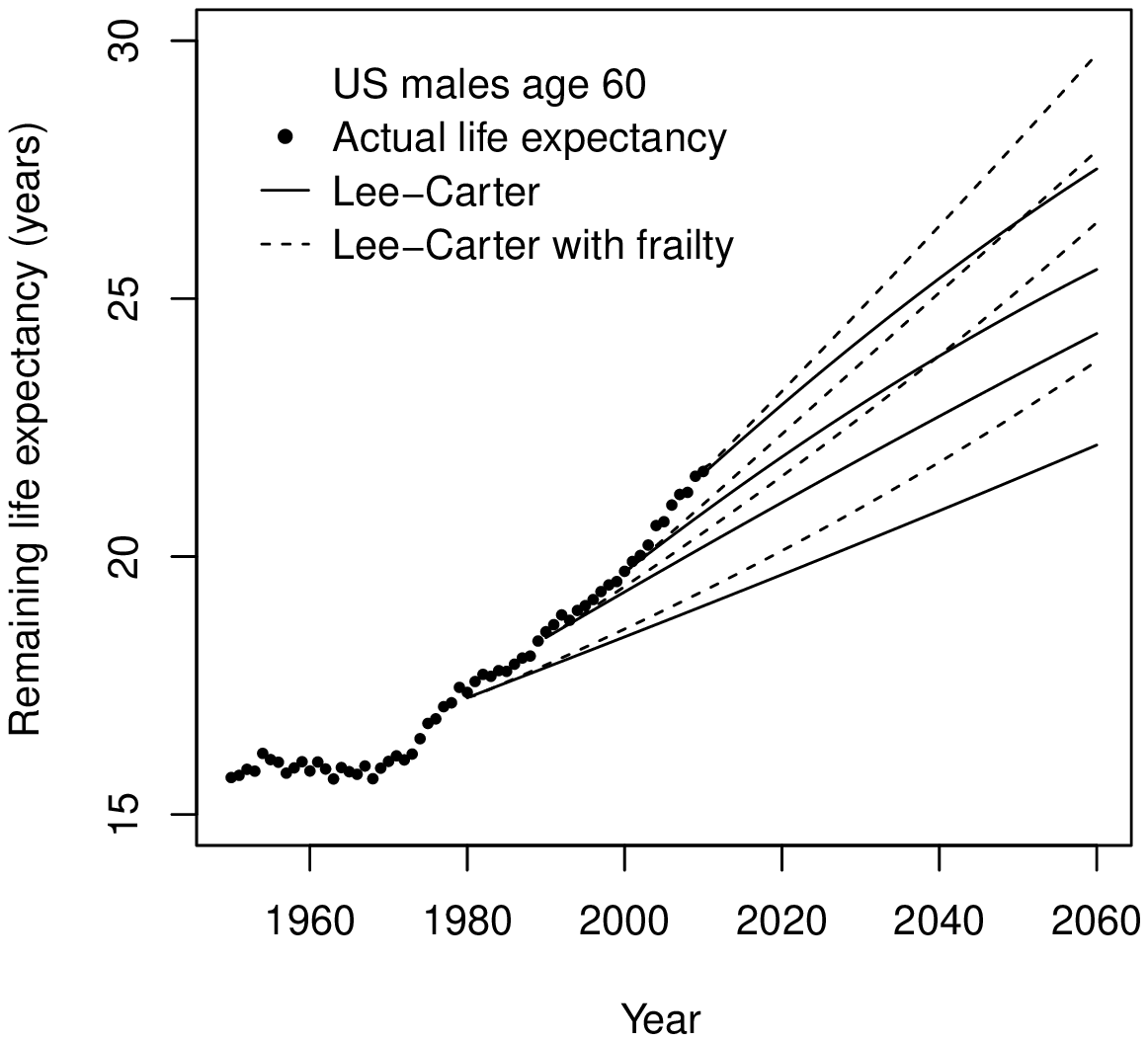}
\end{center}
\vspace*{-5mm}
\caption{Left panel shows estimated (dots) and forecasted index, $k$, of model (\ref{eq:sfmLCapp1})--(\ref{eq:sfmLCapp2}) with $\hat{\sigma}^2=0.73$ and estimation period 1980--2010. 95\%-confidence intervals with and without parameter uncertainty on $\xi$ (drift)
are shown as dashed and solid lines respectively. Right panel shows actual and forecasted life expectancy for US males age 60. Forecasts are based
on model (\ref{eq:sfmLCapp1})--(\ref{eq:sfmLCapp2}) without frailty (solid) and with $\hat{\sigma}^2=0.73$ (dashed) for four different estimation
periods, 1950--1980, 1960--1990, 1970--2000 and 1980--2010.}
\label{fig:LCfrailforecast}
\end{figure}

The left panel of Figure~\ref{fig:LCfrailforecast} shows the estimated index for the estimation period 1980--2010 and frailty variance $\hat{\sigma}^2=0.73$.
The estimated index looks close to linear and a linear forecast seems reasonable. The right panel of Figure~\ref{fig:LCfrailforecast}
shows the corresponding life expectancy forecast for US males age 60 as the dashed line starting in 2010. The forecast is almost linear
while the ordinary Lee-Carter forecast (solid line) curves downwards over time. For comparison the plot also shows Lee-Carter forecasts
with and without frailty based on the estimation periods 1950--1980, 1960--1990 and 1970--2000.\footnote{To show the effect of the estimation period rather than the level of frailty the same value of $\hat{\sigma}^2=0.73$ is used for all forecasts.} The pattern is the same with almost linear frailty forecasts and lower, downward-curving Lee-Carter forecasts.
Note that all historic forecasts, both with and without frailty, are below the actual life expectancy evolution.

\begin{figure}[h]
\begin{center}
\includegraphics[height=7.5cm]{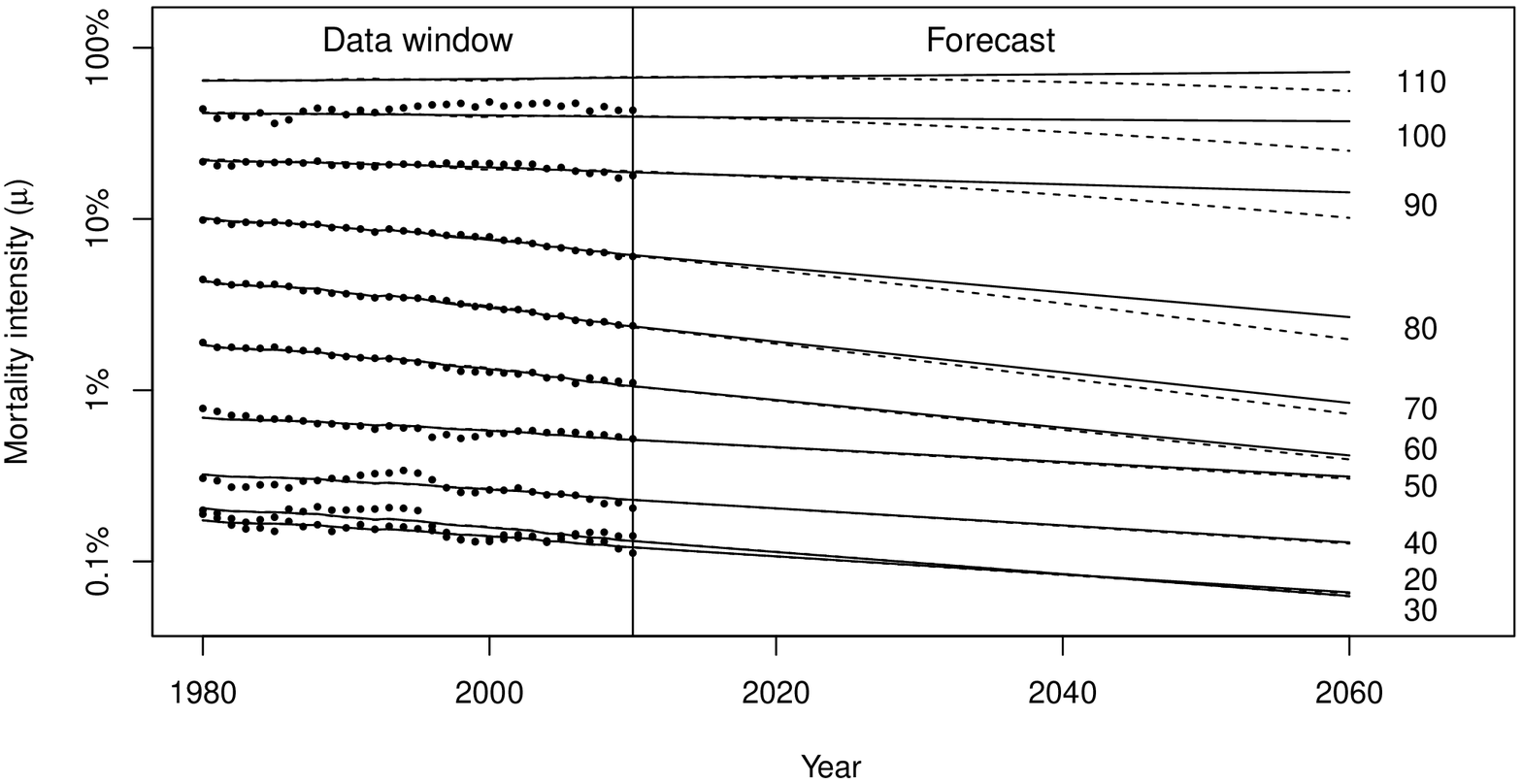}
\end{center}
\vspace*{-5mm}
\caption{The dots show observed death rates for US males from 1980 to 2010 for ages 20, 30, $\ldots$, 100 years.
The fit and forecast of model (\ref{eq:sfmLCapp1})--(\ref{eq:sfmLCapp2}) without frailty (Lee-Carter model) and $\hat{\sigma}^2=0.73$
are shown as solid and dashed lines respectively. The model is fitted to data from 1980--2010 and ages 0--90. Mortality for higher ages
is obtained by logistic extrapolation.}
\label{fig:LCfitfor}
\end{figure}

The effect of frailty in terms of the forecasted age-specific death rates is illustrated in Figure~\ref{fig:LCfitfor}.
For ages below 70 the Lee-Carter forecasts with and without frailty are almost identical. For higher ages
the frailty effect gradually increases and gives rise to the previously announced curving forecasts. In particular,
the frailty forecast predicts substantial improvements in mortality for the very old over the next 50 years while
the ordinary Lee-Carter forecasts predicts essentially no improvements above age 90. The plot also illustrates
that the forecasted age-profile is somewhat distorted. This is a general feature of semi-parametric methods, like the Lee-Carter model,
due to the lack of structure.

History shows that mortality rates generally decline over time, but it also shows changing
rates of improvement. Clearly, there are substantial period effects due to medical breakthroughs,
changes in nutrition and working conditions etc., and frailty can at best be a contributing factor
to the observed increasing rates of improvement. Indeed, the example
shows that adding frailty to historic forecasts is not enough to explain the actual evolution.
However, the addition of frailty leads to projected old-age mortality rates
which at least partly accommodate future higher rates of improvement. The resulting almost linear
life expectancy projections resemble the actual experience better than the downward-curving
Lee-Carter forecasts.

\section{Generalized stochastic frailty models} \label{sec:genSFM}
The theory presented in Section~\ref{sec:SFM} allows for estimation and forecasting of multiplicative frailty models.
This is a useful class of models, but it can also be relevant to consider a larger and more flexible class of additive models
with both frailty and non-frailty components. With additive models we can distinguish between senescent mortality
influenced by frailty and selection and "background" mortality due to e.g.\ accidents with no selection effects. The class introduced below
contains e.g.\ logistic-type models with constant terms which is useful when modeling mortality for a wide age span.

In the following we define the class of generalized stochastic frailty models and we show how to estimate and forecast
these models. Additive models complicates the theory somewhat and we will have to resort to an EM-algorithm for estimation.
The exposition is rather brief and focuses on the areas where
the theory differs from the one presented in Section~\ref{sec:SFM}. Many of the general remarks still apply, but they will not be repeated here.

The models we consider are based on the underlying assumption that the intensity for an individual of age $x$ at time $t$ with frailty $z$
is of the form
\begin{equation}
    \mu(t,x;z) = z\mubase(t,x) + \muback(t,x),
\end{equation}
where $\mubase$ is the baseline intensity influenced by individual frailty and $\muback$ is the background mortality common to all individuals independent of their frailty.
The cohort intensity is then given by
\begin{equation}
   \mu(t,x) = \E[Z|t,x]\mubase(t,x) + \muback(t,x),
\end{equation}
where $\E[Z|t,x]$ denotes the (conditional) mean frailty of the cohort of age $x$ at time $t$, i.e.\ the cohort
born at time $t-x$.

Assume that all cohorts have the same frailty distribution at birth and denote its Laplace transform by $L$.
By a straightforward generalization of the results in Section~\ref{sec:frail} we have $H(t,x) = \nu(I(t,x))$ and
\begin{equation}
   \E[Z|t,x]  = \nu'(I(t,x)) = \nu'\left(\nu^{-1}\left\{H(t,x)\right\}\right),  \label{eq:gmeanfrail}
\end{equation}
where $\nu = -\log L$ and
\begin{align}
    I(t,x) & = \int_0^x \mubase(u+t-x,u)du,  \\
    H(t,x) & = \int_0^x \mu(u+t-x,u) - \muback(u+t-x,u)du.
\end{align}
Note that in contrast to earlier $H$ is not just the integrated cohort intensity. The frailty independent part of mortality, $\muback$, needs
to be subtracted to get the usual connection between $I$ and $H$.

\subsection{Model structure}
Assume we have models for baseline and background mortality of the form
\begin{align}
      \mubase(t,x) & = F(\theta_t,\eta_x),   \label{eq:gmubase}    \\
      \muback(t,x) & = G(\zeta_t,\omega_x),  \label{eq:gmuback}
\end{align}
where $\theta_t$ and $\eta_x$ together with $F$ describe the period and age effects
of frailty dependent mortality, and $\zeta_t$ and $\omega_x$ together with $G$ describe the period and age effects
of frailty independent mortality. All quantities can be multi-dimensional. Further generalizations are possible,
but the chosen structure is sufficient to illustrate the idea.

For ease of presentation we assume that $\eta_x$ and $\omega_x$ are to be estimated from data, but
they may in fact be fixed. As a simple example we could use a Gompertz law for baseline mortality,
$F(\theta_t,x) = \exp(\theta_t^1 + \theta_t^2x)$, and a constant for background mortality, $G(\zeta_t)=\exp(\zeta_t)$.
With no frailty this is the Gompertz-Makeham model, and when frailty is added to baseline mortality
we get the logistic-type model used in Section~\ref{sec:app1}.

A generalized stochastic frailty model is a model where death counts are independent with
\begin{align}
      D(t,x)       & \sim \mbox{Poisson}\left( \mu(t,x) E(t,x)\right),  \label{eq:gPoisson1} \\
      \mu(t,x)     & = \E[Z|t,x]\mubase(t,x) + \muback(t,x),            \label{eq:gPoisson2}
\end{align}
where $\mubase$ and $\muback$ are given by (\ref{eq:gmubase})--(\ref{eq:gmuback}) and $\E[Z|t,x]$ denotes the conditional mean frailty of the cohort of age $x$ at time $t$.
The frailty distribution at birth is the same for all cohorts and it is assumed to belong to a family indexed by $\phi$.
The Laplace transform of the frailty distribution with index $\phi$ is denoted $L_\phi$, and
this is assumed available in explicit form. Further, as a matter of convention we assume that mean frailty is one at birth and we define $\nu_\phi = -\log L_\phi$.

\subsection{Pseudo-likelihood function}  \label{sec:gpseudo}
Estimation will be based on a pseudo-likelihood function in which the problematic term $\E[Z|t,x]$ is replaced
by an estimate. This opens for a general, (relatively) easy estimation procedure which can be implemented
whenever the baseline and background models can be estimated separately.

The idea is to estimate $H$ in the data window by
\begin{equation}
  \tilde{H}(t,x,\zeta,\omega) =  \sum_{u=0}^{x-1} \tilde{m}(u+t-x,u),  \label{eq:gHtilde}
\end{equation}
where
\begin{equation}
   \tilde{m}(t,x) =
   \begin{cases}
   \tilde{m}(t_{\min},x)         & \mbox{for } t < t_{\min} \mbox{ and } x_{\min} \leq x \leq x_{\max}, \\
    m(t,x) - G(\zeta_t,\omega_x)  & \mbox{for } t_{\min} \leq t \leq t_{\max} \mbox{ and } x_{\min} \leq x \leq x_{\max}, \\
   0                             & \mbox{for } 0 \leq x < x_{\min}. \\
   \end{cases}
   \label{eq:gmtilde}
\end{equation}
The reason we need to extend $\tilde{m}$ by (\ref{eq:gmtilde}) is that the summation in (\ref{eq:gHtilde}) falls partly outside the data window.
Apart from the subtraction of background mortality the construction is the same as in Section~\ref{sec:pseudo}.

Appealing to (\ref{eq:gmeanfrail}) we propose to base estimation of model (\ref{eq:gPoisson1})--(\ref{eq:gPoisson2})
on a likelihood function in which $\E[Z|t,x]$ is replaced by $\nu'(\nu^{-1}\{\tilde{H}(t,x,\zeta,\omega)\})$. The resulting approximate likelihood function is referred to as the pseudo-likelihood function,
\begin{align}
    L(\phi,\theta,\eta,\zeta,\omega) & = \prod_{t,x}\frac{\lambda(t,x)^{D(t,x)}}{D(t,x)!}\exp(-\lambda(t,x)),    \label{eq:gpseudoL} \\
    \lambda(t,x) & = \left[ \nu'_\phi(\nu^{-1}_\phi\{\tilde{H}(t,x,\zeta,\omega)\})F(\theta_t,\eta_x)+ G(\zeta_t,\omega_x)\right] E(t,x).
\end{align}
Formally, it corresponds to estimating the modified model
\begin{align}
      D(t,x) & \sim \mbox{Poisson}\left( \mu(t,x) E(t,x)\right), \label{eq:gpseudo1} \\
      \mu(t,x) & = \nu'_\phi(\nu^{-1}_\phi\{\tilde{H}(t,x,\zeta,\omega)\})F(\theta_t,\eta_x) + G(\zeta_t,\omega_x).     \label{eq:gpseudo2}
\end{align}

\subsection{Maximum likelihood estimation} \label{sec:gMLE}
Maximum likelihood estimates of model (\ref{eq:gpseudo1})--(\ref{eq:gpseudo2}) can be obtained by
optimization of the profile log-likelihood function,
\begin{equation}
  l(\phi) = \log L(\phi,\hat{\theta}(\phi),\hat{\eta}(\phi),\hat{\zeta}(\phi),\hat{\omega}(\phi)),
\end{equation}
where $L$ is given by (\ref{eq:gpseudoL}) and $\hat{\theta}(\phi)$, $\hat{\eta}(\phi)$, $\hat{\zeta}(\phi)$ and $\hat{\omega}(\phi)$
denote the maximum likelihood estimates for fixed value of the frailty parameter $\phi$. As argued in Section~\ref{sec:MLest}
the frailty parameter is typically of low dimension, and the profile log-likelihood function can therefore be
optimized by standard optimization routines. The issue remains of how to compute the profile log-likelihood function.

Assume that we have available routines for maximum likelihood estimation of the baseline and background
mortality models separately.
Utilizing that model (\ref{eq:gpseudo1})--(\ref{eq:gpseudo2}) is a special
case of a competing risks model, we can compute the estimates $\hat{\theta}(\phi)$, $\hat{\eta}(\phi)$, $\hat{\zeta}(\phi)$ and $\hat{\omega}(\phi)$
by the following algorithm. The algorithm is an adapted version of the EM-algorithm described in Appendix~\ref{app:EM}.
Let $i=1$ and proceed as follows
\begin{enumerate}
\item Choose initial values for baseline and background parameters, $\theta^0$, $\eta^0$, $\zeta^0$ and $\omega^0$.
\item For all $t$ and $x$ in the data window compute $\tilde{H}(t,x,\zeta^{i-1},\omega^{i-1})$ by (\ref{eq:gHtilde}), and let
$c(t,x) = \nu'_\phi(\nu^{-1}_\phi\{\tilde{H}(t,x,\zeta^{i-1},\omega^{i-1})\})$.
\item Calculate $\theta^i$ and $\eta^i$ as the maximum likelihood estimates for the model
\begin{equation}
           D_{\textrm{base}}(t,x) \sim \mbox{Poisson}\left( c(t,x) E(t,x) F(\theta_t,\eta_x)\right)
\end{equation}
with "data"
\begin{equation}
           D_{\textrm{base}}(t,x) = D(t,x)\frac{c(t,x)F(\theta^{i-1}_t,\eta^{i-1}_x)}{c(t,x)F(\theta^{i-1}_t,\eta^{i-1}_x) + G(\zeta^{i-1}_t,\omega^{i-1}_x)}.
\end{equation}
\item Calculate $\zeta^i$ and $\omega^i$ as the maximum likelihood estimates for the model
\begin{equation}
           D_{\textrm{back}}(t,x) \sim \mbox{Poisson}\left( E(t,x) G(\zeta_t,\omega_x)\right)
\end{equation}
with "data"
\begin{equation}
           D_{\textrm{back}}(t,x) = D(t,x)\frac{G(\zeta^{i-1}_t,\omega^{i-1}_x)}{c(t,x)F(\theta^{i-1}_t,\eta^{i-1}_x) + G(\zeta^{i-1}_t,\omega^{i-1}_x)}.
\end{equation}
\item Increase $i$ by one.
\item Repeat steps 2--5 until convergence.
\end{enumerate}
Note that the two Poisson models are only formal in the sense that $D_{\textrm{base}}$ and $D_{\textrm{back}}$ are not integer-valued, see Appendix~\ref{app:EM} for details.

\subsection{Forecasting}  \label{sec:gforecast}
Forecasting is performed essentially as for the multiplicative case. Assume
that we have maximum likelihood estimates $(\hat{\phi},\hat{\theta},\hat{\eta},\hat{\zeta},\hat{\omega})$ of
model (\ref{eq:gpseudo1})--(\ref{eq:gpseudo2}). Also, assume that we have a forecast, either deterministic or stochastic,
$(\bar{\theta}_t,\bar{\zeta}_t)_{t_{\max} < t \leq t_{\max}+h}$ for a given forecast horizon $h$.
The aim is to forecast baseline, background and cohort mortality for ages $x_{\min} \leq x \leq x_{\max}$ and years
${t_{\max} < t \leq t_{\max}+h}$.

The estimated intensity can be written
\begin{equation}
      \hat{\mu}(t,x)  = \nu'_{\hat{\phi}}(\tilde{I}(t,x))F(\hat{\theta}_t,\hat{\eta}_x) + G(\hat{\zeta}_t,\hat{\omega}_x),
\end{equation}
where $\tilde{I}(t,x) = \nu^{-1}_{\hat{\phi}}(\tilde{H}(t,x,\hat{\zeta},\hat{\omega}))$. Expressing mean
frailty in terms of $\tilde{I}$ facilitates forecasting of cohort mortality as in Section~\ref{sec:fore}.

Baseline and background mortality are forecasted by inserting $\bar{\theta}$ and $\hat{\eta}$ into (\ref{eq:gmubase})
and $\bar{\zeta}$ and $\hat{\omega}$ into (\ref{eq:gmuback}),
\begin{align}
   \mubase(t,x) & = F(\bar{\theta}_t,\hat{\eta}_x), \\
   \muback(t,x) & = G(\bar{\zeta}_t,\hat{\omega}_x),
\end{align}
while cohort mortality is forecasted by
\begin{equation}
   \mu(t,x) = \nu'_{\hat{\phi}}(\tilde{I}(t,x))F(\bar{\theta}_t,\hat{\eta}_x) + G(\bar{\zeta}_t,\hat{\omega}_x),
\end{equation}
where $\tilde{I}$ in the forecast region is given by recursion (\ref{eq:Itildefor}).

Note that in the data window $\tilde{I}$ is given by a transformation of observed death rates with estimated
background mortality subtracted, while in the forecast region it is defined recursively in terms
of the forecasted baseline mortality. In particular, $G$ does not enter $\tilde{I}$ in the forecast.

\section{Final remarks}  \label{sec:remarks}
Populations are without doubt heterogeneous. In this paper we have investigated the use of multiplicative
and additive frailty models taking this fact into account. We have shown how frailty can improve the
fit of simple parametric models, and how it can be combined with the Lee-Carter model to generate
more plausible old-age mortality projections. Obviously, frailty on its own is not enough to
explain the complex dynamics of mortality, but the models can help capture certain essential
features observed in data, e.g.\ changing rates of improvements, otherwise addressed by ad hoc methods.

Stochastic frailty models as here presented offer a general way to combine essentially
any frailty distribution with a parametric or semi-parametric baseline model. This sets
the work apart from the typical use of frailty which rely on matching parametric forms and
closed-form expressions.

The proposed pseudo-likelihood approach is easy to implement for multiplicative models. Additive models
are somewhat harder, but they can be handled by use of the EM-algorithm. In principle, additive models with
more than two components, and multiple sources of frailty, can also be analyzed. However, unless the components target
different segments of data identification might be problematic.

For ease of exposition we have presented the models as single population models. However, it
is often desirable to consider joint models to obtain coherent forecasts for related populations, e.g.\
males and females, subpopulations of a given populations, or similar national populations believed to share
a common trend. A number of models of this type has been proposed, see e.g.\ \citet{lietal04,lilee05,pla09,biacur10,lihar11}.
Apart from notational changes the "fragilization" methodology can be used in this context also.

The models considered are related to the cohort models of e.g.\ \citet{renhab06}
and \citet{caietal09} in the sense that they all focus on the evolution of cohorts through time. The cohort models are designed to capture the so-called
cohort effect seen in some data sets, i.e.\ the phenomenon that the mortality experience of cohorts born in
certain periods differ markedly from the experience of neighboring cohorts, see e.g.\ \citet{wil04,ricetal06}. Cohort effects
could be incorporated in the present framework by allowing the frailty distribution to change over time.

We generally rely on maximum likelihood estimation for determining parameter values. However, if the primary
aim is mortality projection it can be argued that some parameters, e.g.\ frailty parameters,
should not be determined solely on the basis of historic fit but rather on their forecast ability.
We gave an example of how this could be done in the Lee-Carter application, but a more systematic
analysis of how best to determine the frailty distribution remains to be done.

\section*{Acknowledgements}
The material of the paper has been presented in various forms at courses and conferences in recent years.
The comments and feedback from colleagues and students are highly appreciated. In particular,
the author is indebted to Esben Masotti Kryger for numerous stimulating discussions.

\appendix

\section{Estimation of competing risks model} \label{app:EM}
Consider the competing risks model
\begin{equation}
  D(x)  \sim \mbox{Poisson}\left(E(x)[\mu_1(\alpha,x) + \mu_2(\beta,x)]\right), \label{eq:comrisk}
\end{equation}
where $D$ and $E$ denote, respectively, death counts and exposures indexed by age $x$, and $\mu_1$ and $\mu_2$
are mortality intensities with parameters $\alpha$ and $\beta$ respectively. For notational simplicity we do not
include time and we not specify exactly how the parameters enter the two mortality intensities. Apart from
notational changes these omissions are of no consequences for the following discussion.

The interpretation of model (\ref{eq:comrisk}) is that persons of age $x$ can die from two different, independent sources with intensities
$\mu_1$ and $\mu_2$ respectively. The structure is natural to consider in many contexts, but the likelihood function
is complicated and direct estimation of $\alpha$ and $\beta$ can be difficult. This appendix describes how maximum likelihood estimation can
be implemented by the EM-algorithm of \citet{demlairub77} assuming each individual model can be estimated.
The following is a standard application of the EM-algorithm, but since we rely on it in Section~\ref{sec:genSFM} we
here give a brief self-contained exposition for ease of reference.

The EM-algorithm is an algorithm for maximum likelihood estimation for incomplete, or missing, data.
It consists of an E-step in which the expectation of the full log-likelihood is computed
with respect to the missing data, followed by an M-step in which the expectation is maximized
with respect to the parameters. The steps are iterated till convergence.

Imagine in the present setup that deaths had been recorded according to source such that
\begin{align}
  D_1(x) & \sim \mbox{Poisson}\left(E(x)\mu_1(\alpha,x)\right), \label{eq:D1model} \\
  D_2(x) & \sim \mbox{Poisson}\left(E(x)\mu_2(\beta,x)\right),   \label{eq:D2model}
\end{align}
with $D_1(x)$ and $D_2(x)$ independent and $D(x) = D_1(x) + D_2(x)$. Then $D(x)$ would be
distributed according to (\ref{eq:comrisk}) and estimation of $\alpha$ and $\beta$ would be easy.
The full log-likelihood function is given by
\begin{equation}
      l(\alpha,\beta; D_1,D_2)   =     l_1(\alpha;D_1)+    l_2(\beta;D_2),
\end{equation}
where (omitting $x$ for ease of notation)
\begin{align}
   l_1(\alpha;D_1)      & =  \sum_x\{D_1 \log\mu_1(\alpha) - E\mu_1(\alpha)\} + \mbox{constant}, \label{eq:l1} \\
   l_2(\beta;D_2)      & =  \sum_x\{D_2 \log\mu_2(\beta) - E\mu_2(\beta)\} + \mbox{constant}.  \label{eq:l2}
\end{align}
Even though $D_1$ and $D_2$ do not necessarily exist and hence are not "missing" in the normal
sense of the word, we can still use the EM-algorithm based on the missing data interpretation of the model.
The algorithm is as follows:
\begin{description}
\item[] Choose initial values $\alpha_0$ and $\beta_0$ for the parameters.
\item[E-step] Treat $D_1$ and $D_2$ as missing data and compute the expected value
of the full log-likelihood given data $D$ and current parameter estimates
\begin{equation}
   Q(\alpha,\beta) = \E[l(\alpha,\beta;D_1,D_2) | D, \alpha_{i-1}, \beta_{i-1}],
\end{equation}
where
\begin{align}
   D_1 | D,\alpha_{i-1}, \beta_{i-1} & \sim \mbox{Binom}\left(D,\frac{\mu_1(\alpha_{i-1})}{\mu_1(\alpha_{i-1}) + \mu_2(\beta_{i-1})}\right), \\
   D_2 | D,\alpha_{i-1}, \beta_{i-1} & \sim \mbox{Binom}\left(D,\frac{\mu_2(\beta_{i-1})}{\mu_1(\alpha_{i-1}) + \mu_2(\beta_{i-1})}\right).
 \end{align}
 \item[M-step] Maximize $Q$ to obtain new estimates $\alpha_i$ and $\beta_i$.
 \item[] Iterate E-step and M-step until parameter estimates converge.
\end{description}

Both the E-step and the M-step are easy to perform. The E-step merely consists of replacing $D_1$ and $D_2$ in (\ref{eq:l1})--(\ref{eq:l2})
by their conditional expectations, and the M-step consists in estimating the two marginal models. Formally, the EM-algorithm
corresponds to iteratively estimating the models (\ref{eq:D1model})--(\ref{eq:D2model}) with "data"
\begin{align}
  D_1(x) = D(x)\frac{\mu_1(\alpha_{i-1},x)}{\mu_1(\alpha_{i-1},x) + \mu_2(\beta_{i-1},x)}, \\
  D_2(x) = D(x)\frac{\mu_2(\beta_{i-1},x)}{\mu_1(\alpha_{i-1},x) + \mu_2(\beta_{i-1},x)}.
\end{align}
On each iteration "data" are updated according to the latest estimate and the parameters reestimated.
Note, this holds only formally since $D_1$ and $D_2$ computed above are not integer-valued.

It can be shown that the likelihood is increased in each step of the EM-algorithm. The algorithm
hence converges to a local maximum, but it can be rather slow. The advantage is that we can
use the same algorithm to estimate all combinations of models as long as we know how
to estimate the models separately. It is also straightforward to generalize the algorithm
to more than two competing risks.

\bibliographystyle{plainnat}
\bibliography{mortref}

\end{document}